\documentclass[a4paper]{article}

\usepackage[english]{babel}
\usepackage[T1]{fontenc}
\usepackage{textcomp}
\usepackage[utf8]{inputenc}    

\usepackage[a4paper,top=3cm,bottom=2cm,left=3cm,right=3cm,marginparwidth=1.75cm]{geometry}

\usepackage{amsmath}
\usepackage{graphicx}
\usepackage[colorinlistoftodos]{todonotes}
\usepackage[colorlinks=true, allcolors=blue]{hyperref}
\usepackage{svg}
\usepackage{subfig}
\usepackage[separate-uncertainty = true,multi-part-units = repeat]{siunitx}
\usepackage{graphicx,rotating,booktabs}
\usepackage{multirow}
\usepackage[sorting=none, style=nature]{biblatex} 
\usepackage{authblk}

\makeatletter
\newcommand{\settitle}{\@maketitle}
\makeatother

\title{Fractional viscoelastic models for power-law materials}
\author[1]{A Bonfanti \thanks{Corresponding author: Alessandra Bonfanti (ab2425@cam.ac.uk)}}
\author[1]{J L Kaplan}
\author[2,3]{G Charras}
\author[1]{A Kabla \thanks{Corresponding author: Alexandre Kabla (ajk61@cam.ac.uk)}}
\affil[1]{Engineering Department, Cambridge University, UK}
\affil[2]{London Centre for Nanotechnology, University College London, UK}
\affil[3]{Department of Cell and Developmental Biology, University College London, UK }

\begin{document}
\maketitle

\begin{abstract} 
Soft materials often exhibit a distinctive power-law viscoelastic response arising from broad distribution of time-scales present in their complex internal structure. 
A promising tool to accurately describe the rheological behaviour of soft materials is fractional calculus. However, its use in the scientific community remains limited due to the unusual notation and non-trivial properties of fractional operators.
This review aims to provide a clear and accessible description of fractional viscoelastic models for a broad audience, and to demonstrate the ability of these models to deliver a unified approach for the characterisation of power-law materials. The use of a consistent framework for the analysis of rheological data would help classify the empirical behaviours of soft and biological materials, and better understand their response.
\end{abstract}

\section{Introduction}

Characterising and understanding how a material deforms when subjected to external forces is critical for many branches of industry, from food and material processing to product design. For example, in additive manufacturing, solid polymer filaments are melted and then extruded through a nozzle. The material flow behaviour during extrusion affects both the processing time required, and the strength of the final printed object \cite{mackay2018importance,corker20193d}. In the food industry, the texture and mouthfeel of bread is largely dependent on its mechanical properties, which themselves are largely determined by the manufacturing process \cite{tanner2008bread,lazaridou2007effects}. Furthermore, the mechanical response of polymers is often used to infer their composition or microstructure \cite{hata1968effect,van2010decoding}. The scientific study of material deformation is known as \textit{rheology}. To further its inquiry, rheology uses mathematical modelling to capture an approximate representation of the behaviour of materials. This facilitates the classification and comparison of materials, and enables predictions that can then inform engineering design choices and improve manufacturing processes.

Studying how living systems respond to mechanical stimuli is fundamental to gaining a comprehensive understanding of their functions and underlying cellular processes. It is increasingly recognized that the mechanical properties of cells have a great impact on human development and disease progression \cite{park2010measurement,sun2012mechanics,messal2019tissue}. Alterations in the mechanical behaviour of soft tissues when subjected to external stimuli are often associated with diseases \cite{cunningham2005role,phipps2005measurement,moulding2012excess,white2015lung,fritsch2010biomechanical}. Being able to measure and quantify such changes can provide insights into disease evolution, which in turn guide advancements in diagnostic tools and treatments \cite{palacio2015quantitative,remmerbach2009oral,rahimov2013cellular}. In tissue engineering, knowing the response of human tissue to external forces is key to developing replacements that restore the structure and functionality of damaged tissue by matching the properties of the surrounding environment, which promotes cell in-growth and reduces the risk of implant failure \cite{chan2008scaffolding,nimeskern2013quantitative,guilak2014biomechanics,laronda2017bioprosthetic}.

 \begin{figure}[t]
  \centering
 \includegraphics[width=1\textwidth]{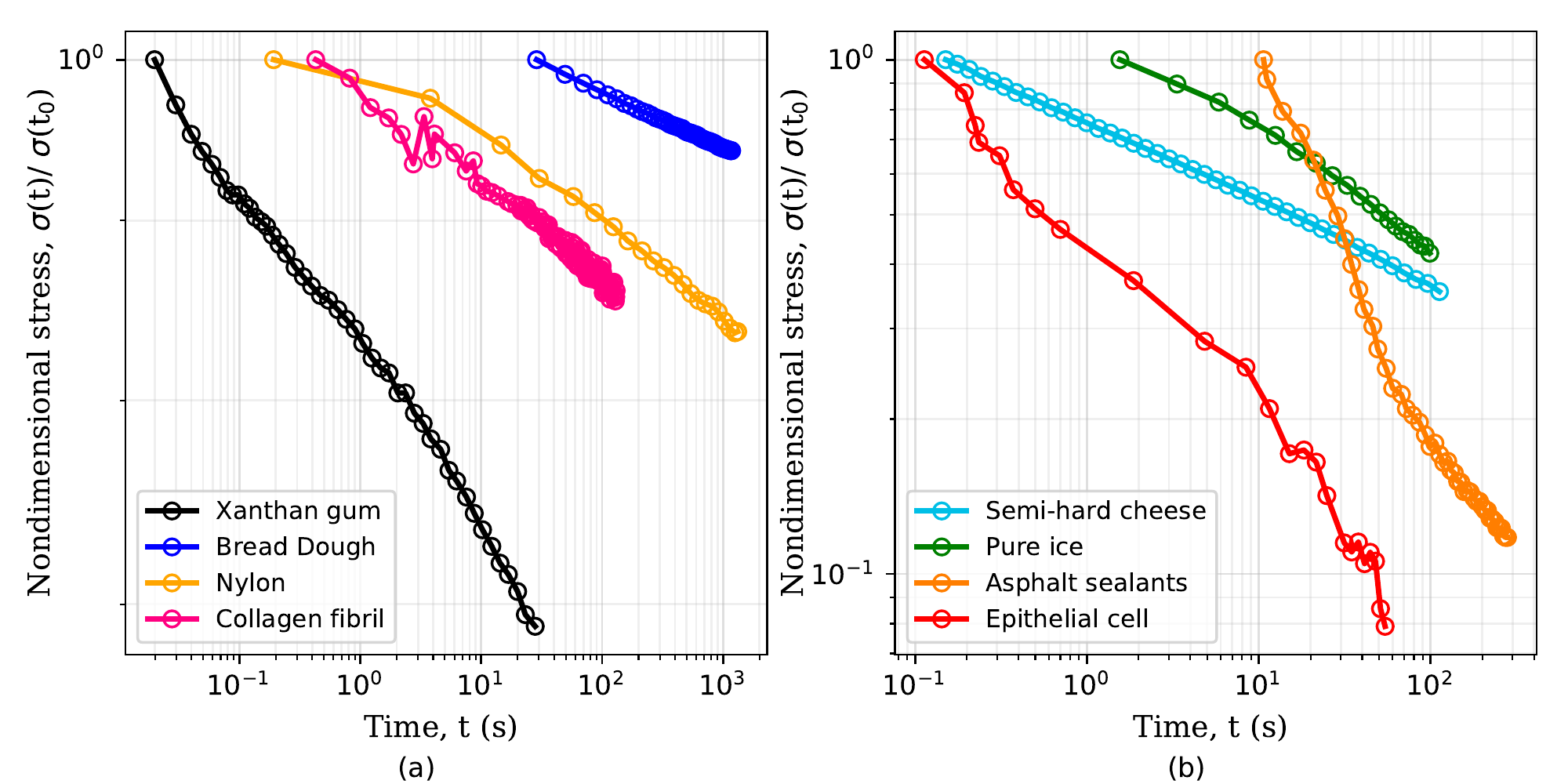}
  \caption{Stress relaxation response of several materials follows a power-law behaviour. Examples of power-law materials include: (a) common polysaccharide (Xanthan gum) used as food additive \cite{gagnon2013undulatory}, bread dough \cite{figueroa2012stress}, synthetic polymers such as nylon of diameter 1.125mm \cite{xu2011stress}, single collagen fibrils \cite{shen2011viscoelastic}; (b) semi-hard zero-fat cheese \cite{faber2017describing}, pure ice at -35$^{\circ}$C \cite{xu2011stress}, asphalt sealants \cite{liu2016preparation} and single MDCK epithelial cell \cite{desprat2006microplates}. For clarity the data has been separated into two subplots and the stresses have been normalized by the initial value for each data set.  }
  \label{fig:02}
\end{figure}

There are two common limits in the behaviour of materials. Elastic materials are solids whose mechanical stress is related to their deformation state with respect to a reference, stress free, geometry; the work done to deform them is entirely stored as recoverable elastic energy. Newtonian fluids are viscous materials whose rate of deformation is function of their mechanical stress; the mechanical work required to deform the material is fully dissipated. Most materials however exist on a spectrum between the elastic and Newtonian limit behaviours; they are referred to as \textit{viscoelastic} materials. A defining characteristic of viscoelastic materials is their time-dependent behaviour. When a constant deformation is applied to viscoelastic materials, their internal stress decreases with time -- this process is known as \textit{relaxation}.
When viscoelastic materials are subjected to a constant load (stress), their deformation (strain) increases with time -- this process is known as \textit{creep}. In applications where long-term durability and shape stability are required, creep may not be a desirable material property. 
In contrast, creep may be a desired property in manufacturing processes where a material is extruded or flows across cavities -- e.g. in additive manufacturing. Various empirical methods are available to quantify the response of viscoelastic materials.
Relaxation tests (in which a constant strain is applied and stress is recorded) and creep tests (in which a constant stress is applied whilst strain is recorded) constitute two testing paradigms for viscoelastic materials. Another common testing methodology requires the application of an \textit{oscillatory load} to the material. 
Typical instruments to collect such data are \textit{Dynamic Mechanical Analysers} (DMA) for materials with solid-like behaviours at long time-scales and \textit{rheometers} for liquid-like materials that need to be studied in shear deformations only. 
Relaxation, creep and oscillatory tests are the three canonical viscoelastic testing paradigms; the most appropriate test can depend on several factors including the experimental hardware available, and the most relevant physical modes of deformation for the application or research question.

Mathematical models are commonly used to describe the creep, relaxation and oscillatory behaviour of viscoelastic materials. One of the primary aims of modelling is to extract model parameters and relate them to the underlying molecular or microstructural deformation mechanisms. In other contexts, such as the food industry, the parameters can be related to sensory perception. Nevertheless, the identification of a suitable model facilitates comparison between the behaviour of different materials, states of organisation or environmental conditions. The theory of linear viscoelasticity is a widely used approach to analyse experimental data that yield to a well-defined mathematical representation of the stress-strain-time relation -- often referred to as \textit{constitutive relationship}. By employing those mathematical methods it is possible to extract parameters that uniquely describe a given material -- often referred to as \textit{material properties} -- and allow the prediction of the evolution of stress and strain within the material under arbitrary loading scenarios.

Advances in mechanical testing hardware have revealed that many materials possessing a complex microstructure exhibit a characteristic power-law signature in their creep, relaxation and spectral behaviours. A selection of data from the literature is shown in figure \ref{fig:02}. Further examples include biological materials -- e.g. tissues \cite{khalilgharibi,serwane2017vivo,nicolle2010strain}, single cells \cite{desprat2005creep,desprat2006microplates,khalilgharibi,fabry2001scaling,darling2006viscoelastic,trepat2004viscoelasticity,ekpenyong2012viscoelastic}, intra and extra cellular components \cite{shen2011viscoelastic,fischer2016rheology,deng2006fast}; gels \cite{meral2010fractional,zhang2018modeling}, polymers \cite{alcoutlabi1998application,xu2011stress}, concrete \cite{bouras2018non,barpi2002creep}, asphalt \cite{blair1944study,mino2016linear,kim1995correspondence}, ice \cite{xu2011stress}, and food--e.g. cheese \cite{subramanian2006linear,faber2017describing}, dough \cite{figueroa2012stress,lefebvre2007pattern}. 
This behaviour represents a challenge to the commonly used viscoelastic models in obtaining a unique mathematical description of the constitutive relation with a limited number of parameters. The power-law behaviour can be approximated with traditional viscoelastic models via a large number of model parameters, which greatly hinders physical interpretation. To circumvent this limitation, traditional models are often discarded in favor of \textit{empirical} fitting functions: mathematical ansatzes derived from qualitative inspection of the data rather than a governing constitutive equation. This approach precludes comparison of parameters across research studies, which can lead to a multitude of different interpretations of similar phenomena \cite{wu2018comparison}.

A promising solution relies on the use of fractional calculus, a branch of mathematics that extends integration and differentiation operators to non-integer order \cite{rudolf2000applications,magin2006fractional}, to enrich the classical linear viscoelastic framework \cite{blair1947limitations,blair1947role}. This led to the development of a new formalism for the modelling of viscoelastic materials known as fractional viscoelasticity. Fractional viscoelasticity has been applied to complex geological and construction materials such as bitumen (asphalt)~\cite{blair1944study,mino2016linear}, concrete~\cite{barpi2002creep,bouras2018non}, rock mass~\cite{zhou2011creep,wu2015improved,ding2017unexpected,liao2017fractional,zhang2018long,ma2018new}, waxy crude oil~\cite{zhang2016rheological,hou2014new}, as well as polymers and gels~\cite{alcoutlabi1998application,chen2004dynamic, bouzid2018computing, aime2018power, kaplan2019pectin}, and food \cite{faber2017describing}. Numerous examples can also be found of fractional viscoelasticity applied to biological materials such as epithelial cells~\cite{bonfanti2019unified}, breast tissue cells~\cite{coussot2009fractional,carmichael2015fractional}, lung parenchyma~\cite{dai2015model}, blood flow~\cite{perdikaris2014fractional,yu2016fractional}, as well as red blood cell membranes \cite{craiem2010fractional}. In spite of the examples reported above, when we consider the number of power-law responses observed in the literature (some of which are modelled empirically), fractional viscoelasticity remains significantly underused. By using a consistent formalism for the analysis of power-law materials, a direct comparison of parameters across studies can be achieved \cite{bonfanti2019unified}, which greatly extends the use of available data.

The main aims of this review are to demonstrate the applicability of fractional viscoelasticity to a wide range of real problems, and to carefully illustrate its advantages over commonly used modelling approaches. Furthermore, we aim to present fractional viscoelasticity in a way that is accessible to researchers without specialised knowledge of fractional calculus. To achieve this, we first illustrate the limitations of spring-dashpot models in capturing power-law viscoelasticity. Following this, the mathematical foundations of fractional viscoelasticity are introduced. 
We will describe the behaviour of networks of fractional viscoelastic elements with a focus on their limit behaviours are short and long time-scales. 
A number of studies which originally used empirical or spring-dashpot based traditional models are then revisited in order to demonstrate the benefits of fractional viscoelastic models. A physical interpretation of the phenomena is discussed where possible.

\section{Introduction to linear viscoelasticity}

The linear theory of viscoelasticity provides a powerful mathematical framework to link stress, strain and time and provide predictions of stress and strain distribution during arbitrary loading conditions. 
A material is assumed linear viscoelastic if the stress function $\sigma(t)$ and strain function $\epsilon(t)$ are linearly connected: 
(i) if the strain function $\epsilon(t)$ is multiplied by a constant factor, the resulting stress would be scaled by the same constant, and vice-versa.
(ii) the response to a strain (or stress) that is a linear combination of two arbitrary strain (or stress) functions is given by the same linear combination of their two individual responses.

Most materials exhibit linear, or quasi linear, behaviour for small deformations while their response becomes nonlinear at large deformations. The linear theory of viscoelasticity is a commonly used approximation for the study of materials' behaviour as it leads to an accessible and manageable mathematical formulation of the stress, strain and time relationship. Furthermore, the theory of linear viscoelasticity represents a good starting point for the study of more complex nonlinear responses that often gives rise to more complex models \cite{findley2013creep, keshavarz2017nonlinear,bharadwaj2017strain,muller2011nonlinear}.

A common test performed to characterise a viscoelastic material is to analyse the stress response of the material when subjected to a constant strain (relaxation test).
The implication of the linearity assumption is that the resultant stress function scales \textit{linearly} with the magnitude of the step in strain~\cite{kailath1980linear}. Thus, for a step in strain of amplitude $\epsilon_0$ at time $t=0$, the resulting stress $\sigma(t)$ can be written
\begin{equation}\label{eq:1.1}
  \sigma(t) =  G(t) \; \epsilon_0,
\end{equation}
where $G(t)$ is the relaxation modulus, a monotonically decreasing function. Similarly, in the case of a creep test, the strain response $\epsilon(t)$ is proportional to the stress step amplitude $\sigma_0$ that is imposed at $t=0$:
\begin{equation}\label{eq:1.2}
  \epsilon(t) =  J(t) \; \sigma_0,
\end{equation}
where $J(t)$ is the creep modulus, a monotonically increasing function of time.

A key consequence of the linearity assumption is that the response of the material at time $t$ is given by the sum of the responses to the perturbations imposed at all previous times. Therefore, given an arbitrary stress or strain history, we can compute either the strain response or stress response respectively by integrating the entire history of infinitesimal `step' excitations~\cite{gutierrez2014constitutive,lakes2009viscoelastic}. Assuming that the material does not age, i.e. that its mechanical behaviour does not change with time during the course of a measurement, the strain (or stress) response at time $t$ of a step in stress (or strain) at time $\tau$ would be given by the relaxation function $G(t-\tau)$ (or creep function $J(t-\tau)$). This results in the following convolution integrals:
\begin{equation}\label{eq:1.3}
	\sigma(t) =  \int_{0}^{t} G(t-\tau) \frac{d\epsilon(\tau)}{d\tau} d\tau,
\end{equation}
\begin{equation}\label{eq:1.4}
	\epsilon(t) =  \int_{0}^{t} J(t-\tau) \frac{d\sigma(\tau)}{d\tau} d\tau,
\end{equation}

Another consequence of linearity is that, when a viscoelastic material is subjected to a sinusoidal stress, the recorded strain has the same frequency, but with a phase difference that may depend on frequency. For purely elastic materials the phase difference is \ang{0} and for purely viscous materials the phase difference is \ang{90}. For viscoelastic materials, the phase difference lies between these two values. If we assume a linear material, by considering an oscillatory excitation $\epsilon(t) = e^{i \omega t}$, we obtain a stress response $\sigma(t) = G^{*} e^{i \omega t}$~\cite{lakes2009viscoelastic} from which we can define the so-called \textit{complex modulus}, or \textit{dynamic modulus}:
\begin{equation}
   G^{*}(\omega) = G^{\prime}(\omega) + i G^{\prime \prime}(\omega),
\end{equation}
where $\omega$ is the frequency. The real part of this complex stress response, $G^{\prime}(\omega)$, is defined as the storage modulus (as the energy is stored in an ideal elastic material). The imaginary part of the response, $G^{\prime \prime}$, is defined as the loss modulus (as energy is dissipated in a purely viscous material). 

For a linear material, the three moduli, relaxation, creep and dynamic moduli, are directly related. Their relationship can be expressed in the Laplace domain. For instance, $\widetilde{G}(s)\widetilde{J}(s) = s^{-2}$, where $\widetilde{G}(s)$ and $\widetilde{J}(s)$ are the Laplace transforms of equations \eqref{eq:1.3} and \eqref{eq:1.4} respectively \cite{findley2013creep}. Therefore, we can predict one type behavioural mode (e.g. creep) from observing a different behavioural mode (e.g. relaxation).

\section{Characteristics of traditional linear viscoelastic models}

\begin{figure}[t]
 \centering
 \includegraphics[width=1.0\textwidth]{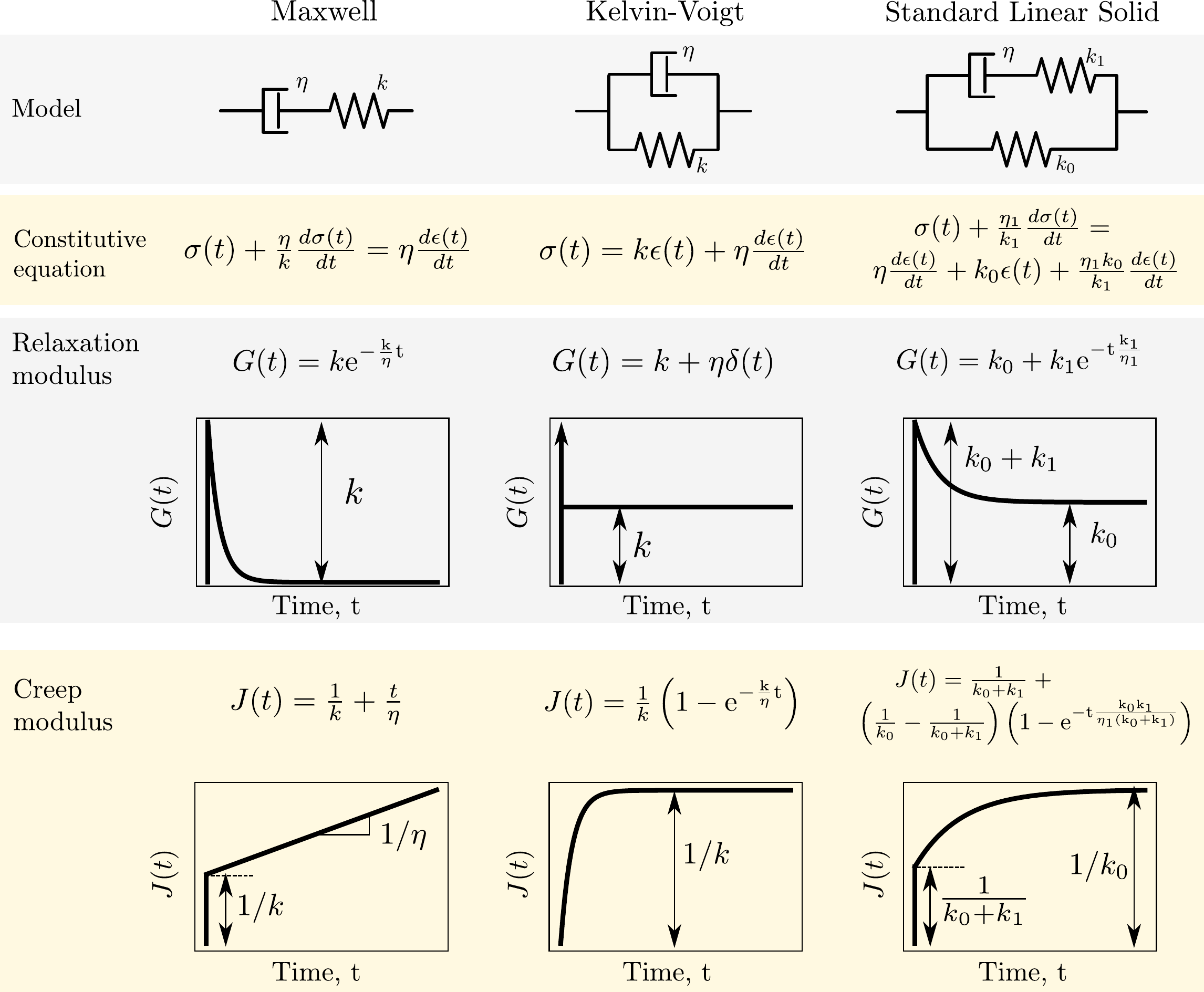}
 \caption{Properties of the three common viscoelastic material models.}
 \label{fig:04}
\end{figure}

The expressions above are generic and applicable to any material that behaves in a linear manner, being solid and liquid like. Specific expressions for the moduli -- i.e. relaxation, creep and dynamic moduli -- can be deduced in several ways. A common approach relies on the use of fundamental viscoelastic units that can be combined in series or parallel -- this element-by-element model building process is analogous to electrical circuit models. The most commonly used viscoelastic units are the Hookean spring and the Newtonian dashpot, with governing equations of $\sigma(t) = k \epsilon(t)$ and $\sigma(t) = \eta \dot{\epsilon}(t)$ respectively. The simplest models that can be formed from these two viscoelastic units are their series (Maxwell model) and parallel (Kelvin-Voigt model) combinations, both of which are shown in figure~\ref{fig:04}. The creep and relaxation moduli can be derived from the corresponding differential equation by: a) assuming that the stress is constant and solving for the differential equation in strain, leading to the creep modulus, and b) assuming that the strain is constant and solving for the differential equation in stress, leading to the relaxation modulus. The complex modulus is derived from the constitutive differential equation, either by considering a oscillatory forms of the stress and strain functions, or by taking the Fourier transform of the differential equation, leading to an algebraic equation which can be rearranged for $G^{*} = \hat{\sigma}(\omega)/\hat{\epsilon}(\omega)$, where the hat ( $\hat{}$ ) symbol denotes a Fourier transformed function.

\begin{figure}[t]
    \begin{minipage}{0.40\columnwidth}
    \subfloat[]{\includegraphics[height=9cm]{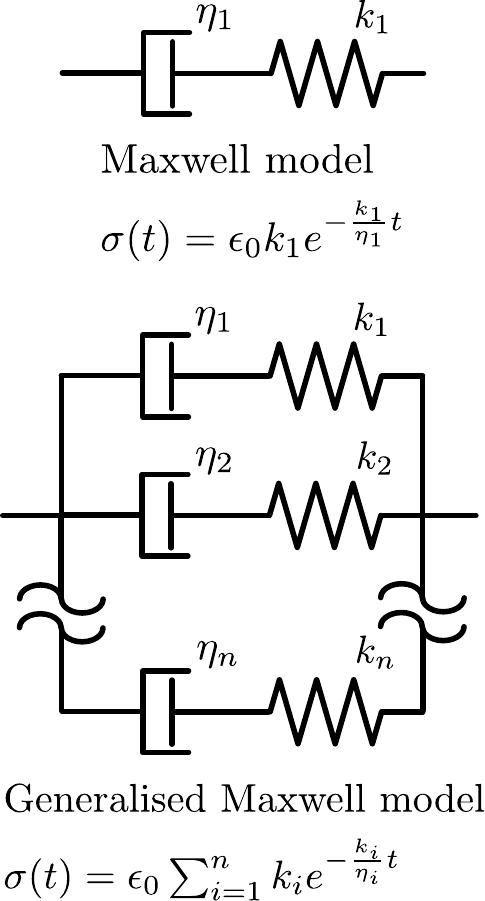}}
    \end{minipage}
    \begin{minipage}{0.48\columnwidth}
    \subfloat[]{\includegraphics[height=5cm]{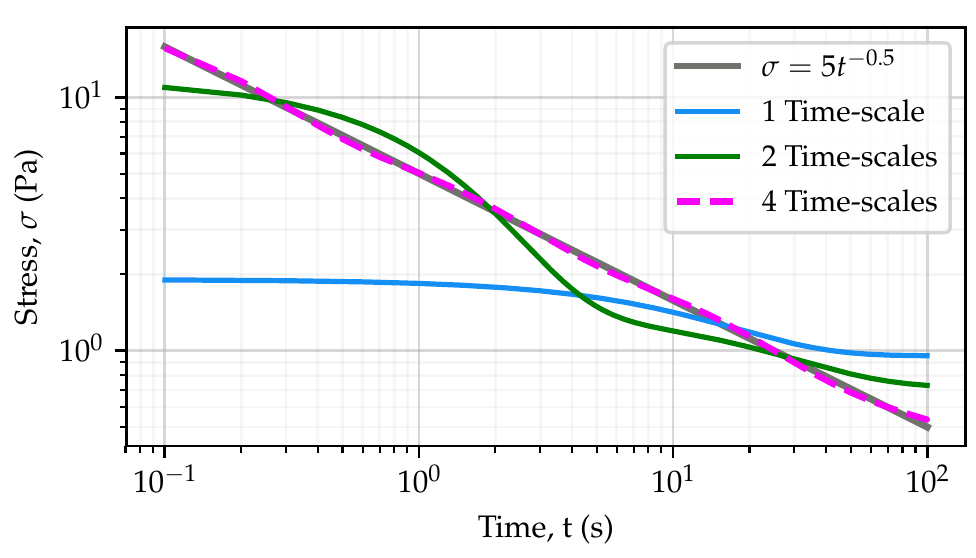}}
    \\[3mm]
    \subfloat[]{\includegraphics[height=5cm]{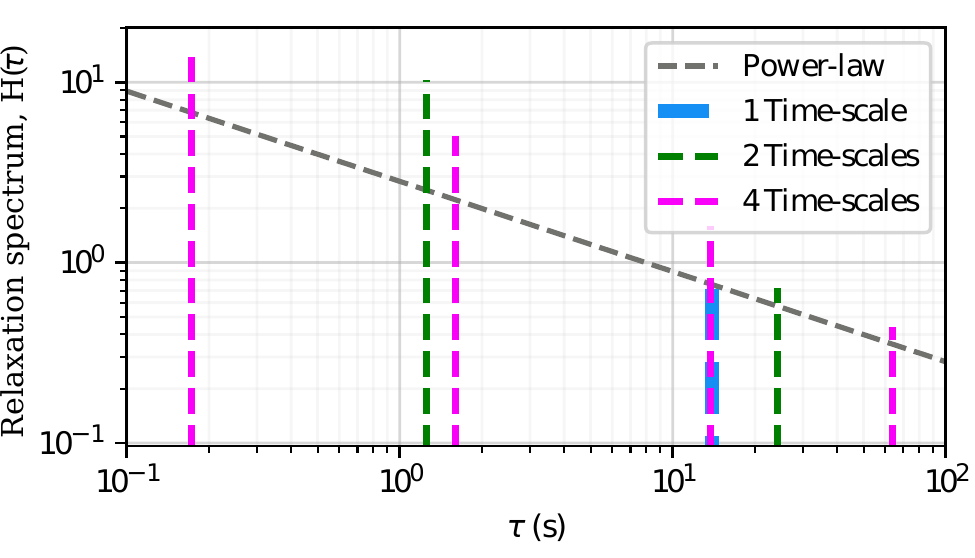}}
    \end{minipage}
    \caption{Use of traditional viscoelastic models to analyze power-law responses. (a) Sketch of the Maxwell model and its generalised form with the correspondent relaxation modulus showing an exponential form. (b) Approximation of a power-law response $\sigma(t) = 5\cdot t^{-0.5}$ (grey line) using the generalised Maxwell model with $n=$ 1 (blue), 2 (green), 4 (dashed purple), where for $n=1$ it corresponds to the Maxwell model. Fitted model parameters in table \ref{tab:0} in Appendix. (c) Relaxation spectra of the power-law response and the generalised Maxwell model in with $n=$ 1, 2, and 4 (b).}
      \label{fig:05}
\end{figure}

As the differential equations relating stress and strain of traditional spring-dashpot models take the form of a linear ordinary differential equation with constant coefficients, both relaxation and creep moduli involve exponential functions (see figure \ref{fig:04}). As an example of this, consider the stress response of a Maxwell model to a step in strain: 
 \begin{equation}\label{eq:1bis}
  \sigma(t) =  k \rm{e}^{-t\frac{k}{\eta}} \epsilon_0 = G(t) \epsilon_0,
\end{equation}
where $G(t)$ is the relaxation modulus. The ratio $\tau = \eta/k$ has the units of time and can be considered as the characteristic relaxation time of the material. This time-scale is a useful quantity; it represents the time required for the stress to fall to $1/e$ of its initial value. More importantly, the time-scale may describe a physical relaxation process more effectively than the values of $\eta$ and $k$ separately, such as rates of binding or unbinding of molecular structures. To capture more complex material behaviours, additional elastic springs and viscous dashpot elements may be used. These additional elements give rise to multiple time-scales. Some models, for example the generalised Maxwell model shown in figure \ref{fig:05} (a), give rise to an arbitrarily large number of time-scales. (See \cite{roylance2001engineering,findley2013creep,bland2016theory} for more details about spring-dashpot based linear viscoelastic models.)


Power-law viscoelastic materials correspond to systems with a broad range of relaxation/creep time-scales arising from dissipation/deformation modes occurring at different time and length scales. Mathematically, power-law behaviour consists not of a discrete number of time-scales, but a continuous distribution of time-scales. For this reason, power-law rheology can only be \textit{approximated} by the spring-dashpot models that yield exponential terms with discrete time-scales~\cite{svensson1973approximation,beylkin2005approximation,winter1986analysis}. The larger the number of exponential terms, the closer the approximation~\cite{halldin2014constitutive,bembey2006viscoelastic,efremov2017measuring,wang2016modeling}. We demonstrate this in figure~\ref{fig:05} by incrementally increasing the number of Maxwell arms in a Generalised Maxwell model, each of which contributes a material time-scale. With four Maxwell arms, a good approximation of a power-law is obtained over the time domain of the data (figure \ref{fig:05} (b)). Another informative perspective on the approximation process can be gained via \textit{relaxation spectrum}, which represents the relative behavioural dominance at different times. The relaxation spectrum found here for the power-law relaxation modulus was approximated as~\cite{christensen2012theory,lakes2009viscoelastic}
\begin{equation}
    H(\tau) =- \left. \frac{dG(t)}{d\ln{t}} \right|_{t = \tau}
\end{equation}
where $G(t)$ is the relaxation modulus. Figure \ref{fig:05} (c) shows the relaxation spectrum of the power-law material in logarithmic scale, which is a power-law distribution of time-scales with the same exponent as the original relaxation data. The optimal fit of the various tested Maxwell models can be seen to yield time-scales that are equally spaced in logarithmic scale. As more Maxwell arms are added, a broader coverage of the true power-law distribution is attained. 

This qualitative assessment of the above relaxation responses and their corresponding spectra shows that spring-dashpot models might be sufficient to approximate power-law behaviour. However, the approach has a number of disadvantages. A large number of model parameters make the computational analysis far more expensive, and interpretation of their physical meaning becomes significantly more difficult. The characteristic times extracted from the fitting process mostly capture the time window of the data and do not represent intrinsic material properties. Moreover, the fitting would optimise the match with data in the region fitted, but provide poor predictions at time-scales shorter of longer than those provide in the original data.

To circumvent the above disadvantages inherent to spring-dashpot based approximations, empirical power-law expressions have been used as a simple way to capture experimental measurements \cite{desprat2005creep,fabry2001scaling,kobayashi2017simple,bonakdar2016mechanical,khalilgharibi}. These expressions are essentially a mathematical ansatz, formulated to capture the main qualitative features of the experimental data. The end result of this formulation is usually a single modulus of interest that may or may not be analytically converted into the other moduli, necessitating numerical analysis for the prediction of the material response to other external loading patterns. Further, the use of such \textit{ad-hoc} models limits the scope of the measurements since the model parameters may not be easily comparable across studies.

\section{Fractional viscoelastic models}

With regards to the modelling of power-law viscoelastic materials, we have now outlined the advantages and disadvantages of the phenomenological spring-dashpot approach and the empirical \textit{ad-hoc} approach. In this section we introduce an additional viscoelastic element, the \textit{springpot}, which is able to capture a power-law behaviour with a minimum number of parameters. We then demonstrate how the springpot can be combined with other elements in series and parallel configurations to capture complex power-law signatures in various contexts. 

\begin{figure}[t]
  \centering
 \includegraphics[height=4cm]{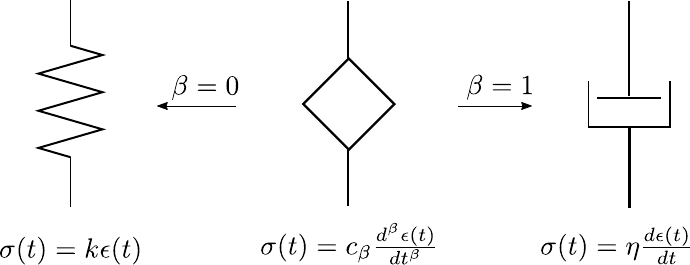}
  \caption{Sketch of the fractional element--springpot. It behaves as a spring when $\beta = 0$ and as a dashpot when $\beta = 1$. }
  \label{fig:f2}
\end{figure}

\subsection{The springpot captures power-law behaviour}

By inspecting the power-law relaxation data (figure \ref{fig:02}), it can be seen that the relaxation modulus takes the form $G(t) = A t^{-\beta}$, where $A$ is a constant and $\beta$ the power exponent. Substituting this into equation~\ref{eq:1.3}, the relationship between stress and strain for a power-law material becomes:
\begin{equation}\label{eq:2}
	\sigma(t) =  A \int_{0}^{t} (t-\tau)^{-\beta} \frac{d\epsilon(\tau)}{d\tau} d\tau.
\end{equation}
A branch of Mathematics called Fractional Calculus provides the definition of the generalization of the differentiation operation to non-integer order valid for a function $f(t)=0$ for $t<0$ as \cite{oldham1974fractional}
\begin{equation}\label{eq:2_1}
	\frac{d^\beta f(t)}{dt^\beta} = \frac{1}{\Gamma(1-\beta)} \int_{0}^{t} (t-\tau)^{-\beta} \frac{df(\tau)}{d\tau} d\tau,
\end{equation}
where $\Gamma(\cdot)$ is the gamma function. Redefining the coefficient $A=c_\beta/\Gamma(1-\beta)$ in equation \ref{eq:2} yields the following result:
\begin{equation}\label{eq:2_2}
	\sigma(t) = \frac{c_\beta}{\Gamma(1-\beta)} \int_{0}^{t} (t-\tau)^{-\beta} \frac{d\epsilon(\tau)}{d\tau} d\tau = c_{\beta} \frac{d^\beta \epsilon(t)}{dt^\beta},
\end{equation}
which provides a simple constitutive equation for power-law materials. A more detailed derivation of this relationship is presented in the next section. 

The springpot's schematic symbol and relationships to the spring and dashpot elements are shown in figure~\ref{fig:f2}. For dimensional consistency, the unit of the constant $c_{\beta}$ must be Pa~s$^{-\beta}$. Due to its unusual physical unit, the parameter $c_\beta$ lacks a tangible physical meaning, but is often interpreted as the `firmness' of a material \cite{blair1942subjective}. 
It should be noted that in the limiting cases where $\beta = 0$ or $\beta = 1$, the springpot reduces to a spring or dashpot respectively, and the constant $c_{\beta}$ represents the elastic spring constant, $k$ (Pa), or the dashpot viscosity, $\eta$ (Pa s), respectively.

To overcome the use of a constant $c_{\beta}$ whose physical meaning may be unclear, the constitutive equation may be written as $\sigma(t) = \lambda_0 \tau_0^\beta d^\beta \epsilon(t)/dt^\beta$, where $\lambda_0$ has units of Pa and $\tau_0$ is a characteristic time that has units of s \cite{schiessel1995generalized,mainardi2011creep}. 
However, it is not possible to design experiments to separately measure $\lambda_0$ and $\tau_0$. Therefore, in practical problems, it is more relevant to define the fractional element with two parameters $c_\beta$ and $\beta$ \cite{jaishankar2013power}, and this is what we use in the current review.

\subsubsection{Derivation of the springpot's governing equation}
Although not essential for springpot practitioners, it is illuminating to derive the springpot's governing equation. 
Whilst the correspondence between power-law viscoelasticity and the fractional derivative has been presented before \cite{podlubny1998fractional,surguladze2002certain}, here we attempt summarize the main steps of the derivation in a pedagogically accessible way.

Fractional derivatives can be understood as generalised integrals. Integrating a function (here $\epsilon(t)$) an integer $n$ number of times, we get:
\begin{equation}\label{eq:3bis}
	\left(I^n f\right) (t) =  \int_{0}^{t} \dots \left[ \int_{0}^{\tau_2} \left( \int_{0}^{\tau_1} f(\tau_0) d{\tau_0} \right) d{\tau_1}\right] \dots d{\tau_{n-1}}  = \frac{1}{(n-1)!} \int_{0}^{t} (t-\tau)^{n-1} f(\tau) d{\tau},
\end{equation}

where the final simplification on the RHS made use of Cauchy's repeated integral formula. Equation~\ref{eq:3bis} can in fact be generalised to any positive real number $\alpha$ by recalling the extension of factorial numbers to non-integer values via the Gamma function $\Gamma(\alpha) = (\alpha-1)!$
\begin{equation}\label{eq:5}
	\left(I^\alpha f\right) (t) = \frac{1}{\Gamma(\alpha)} \int_{0}^{t} (t-\tau)^{\alpha-1} f(\tau) d{\tau} .
\end{equation}
If we substitute $\alpha = 1 - \beta$ into equation \eqref{eq:5} and $f(t) = \frac{d{\epsilon(\tau)}}{d{\tau}} $, we can then re-write the expression for the total stress reported in equation \eqref{eq:2} as follows
\begin{equation}\label{eq:6}
	\sigma(t) =  \frac{c_\beta}{\Gamma( 1-\beta)} \int_{0}^{t} (t-\tau)^{-\beta} \frac{d{\epsilon(\tau)}}{d{\tau}} d{\tau} = c_\beta \left(I^{1-\beta} \frac{d{\epsilon}}{d{t}}\right) (t).
\end{equation}
Since the fundamental relation $(I^{a+b} f)(t) = I^a\left(I^b f\right)(t)$ holds \cite{oldham1974fractional}, equation \eqref{eq:6} can be expressed as
 \begin{equation}\label{eq:7}
	\sigma(t) =  c_\beta I^{-\beta} \left(I^1\frac{d{\epsilon}}{d{t}}\right) (t) = c_\beta \left(I^{-\beta} \epsilon\right)(t)
\end{equation}
where $\left(I^{-\beta} \epsilon\right)(t)$ is the definition of fractional derivative $\left(D^{\beta} \epsilon\right)(t)$. Equivalent to the above, we can write
\begin{equation}\label{eq:9}
	\sigma(t) = c_{\beta} \frac{d^\beta \epsilon(t)}{dt^\beta},
\end{equation}
which is the governing equation of the springpot~\cite{blair1947role}.

It should be noted that in the literature, different definitions of the fractional derivative have been put forward. Here we use the Caputo's derivative definition since it naturally arises from experimental observations and it has shown to have a better applicability to real problems, where initial conditions are known in terms of derivatives of integer index \cite{di2011visco}. If we assume that the system is at rest for time $t\leq0$, the fractional derivative is given by \cite{mainardi2008} 
\cite{mainardi2008} 
\begin{equation}\label{eq:8}
	^{C}D^{\beta} f(t) = \frac{d^\beta f(t)}{d{t}^\beta} = \frac{1}{\Gamma(1-\beta)} \int_{0}^{t} (t-\overline{t})^{-\beta} \frac{df(\overline{t})}{d\overline{t}} d\overline{t},
\end{equation} 
where $0<\beta<1$ and $C$ denotes a Caputo fractional derivative.. So to obtain the value of the Caputo fractional derivative at time $t$, we must integrate from the initial time $t=0$. In other words the fractional derivative is a \textit{non-local operator}. This is in contrast to integer order derivatives, whose value is determined by the limiting behaviour of the function at the evaluated point. Although hysteresis is also present in the generic hereditary integral shown in equation~\ref{eq:1.3}, it is a fundamental feature of the fractional derivative and consequently the springpot. Lastly, it is important to note that the fractional operator is a linear operator. Therefore, the springpot element lies within the framework of linear viscoelasticity.

\subsubsection{Rheological behaviour of the springpot}

\begin{figure}[t]
  \centering
 \includegraphics[width=1\textwidth]{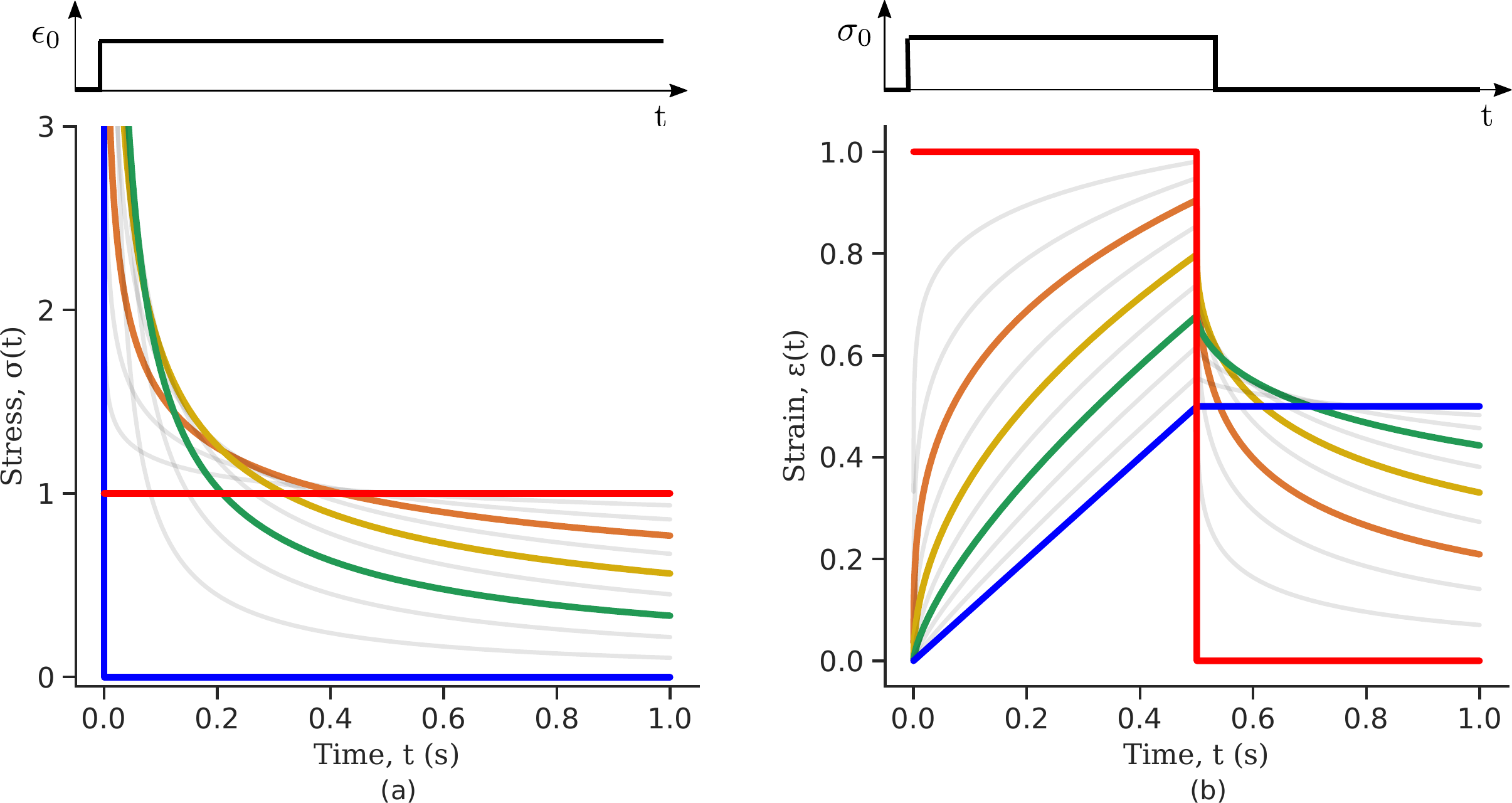}
  \caption{Responses of a springpot when subjected to (a) constant strain and (b) a step in stress. The red curves are the responses for a spring ($\beta =0$), the blue curves for a dashpot ($\beta =1$) and the grey curves are for increasing values of $\beta$ from 0.1 to 0.9 (orange, yellow and green curves are respectively $\beta =$ 0.3, 0.5 and 0.7). }
  \label{fig:f3}
\end{figure}

Despite the simple constitutive equation of the fractional springpot element, it possesses rich behavioural diversity. It is a generalization of the classical viscoelastic elements, the spring and the dashpot, and exhibits a behaviour that is intermediate between the two (see figure \ref{fig:f2}). 
By substituting a step of deformation $\epsilon(t) = \epsilon_0 H(t)$ (where $H(t)$ is the Heaviside step function) into equation \eqref{eq:9} and making use of the definition of the fractional derivative in equation \eqref{eq:8}, we can extract the relaxation modulus $G(t)$ of the springpot \cite{di2011visco}:
\begin{equation}\label{eq:3}
	\sigma(t) = G(t) \:\epsilon_0 \text{ where } G(t) = \frac{c_{\beta} }{\Gamma(1-\beta)} t^{-\beta}
\end{equation}
The relaxation modulus is as expected a power-law function of the time whose exponent matches the order of the fractional derivative. As a result, the corresponding relaxation spectrum of the springpot is also a simple power-law of exponent $\beta$, and therefore captures very well, by design, the behaviour of power-law materials, with only two parameters.
The creep modulus $J(t)$ can be deduced from the expression of $G(t)$. Both are related in the Laplace domain by $\widetilde{G}(s)\widetilde{J}(s) = s^{-2}$. The Laplace transform of $G(t)$ in equation \eqref{eq:3} is given by $\widetilde{G}(s) = c_\beta s^{\beta-1}$. Therefore, $\widetilde{J}(s) = 1/(c_\beta s^{1+\beta})$, whose inverse Laplace gives the creep modulus \cite{di2011visco}
\begin{equation}\label{eq:6bis}
	J(t) = \frac{1}{c_\beta \Gamma(1+\beta)}t^\beta.
\end{equation}

Qualitatively, the linear elastic solid and the Newtonian fluid behaviours are the two limit behaviours of the springpot for $\beta$ = 0 and 1 respectively. To understand the interpolation of the springpot between elastic and viscous behaviour, we plot the springpot response to a step in strain for different values of $\beta$ (figure \ref{fig:f3} (a)). When a dashpot ($\beta=1$) is subjected to a step in strain, the stress is initially infinite but immediately dissipated afterwards (figure \ref{fig:f3} (a), blue curve). By gradually decreasing the exponent $\beta$ the time taken by the springpot to dissipate the stress increases until the limit of the linear elastic solid is reached. When a spring ($\beta=0$) is subjected to a constant deformation, the stress reached is proportional to the spring constant $k$ (figure \ref{fig:f3} (a), red curve).

To illustrate the creep behaviour of the springpot, we consider its response to a step in stress $\sigma_0$ imposed for a limited time $t^*$ -- loading phase -- and then returned to 0 -- unloading phase. Figure \ref{fig:f3} (b) shows plots of the system's response for different values of $\beta$ during the loading and unloading phases. 
A spring generates a strain directly proportional to the stress, as expected, and it immediately returns to its original length upon unloading. A dashpot linearly deforms while the stress is imposed, and does not return to its initial state after unloading.
The response of the springpot is more complex.
By applying the superposition principle, we can write an expression for the springpot response ($0<\beta<1$) after the load is removed at time $t = t^*$ that is given by
\begin{equation}\label{eq:10}
	\epsilon(t) =  \frac{\sigma_0}{c_\beta \Gamma(\beta)}(t^\beta-(t-t^*)^\beta),
\end{equation}
noting that when $t$ tends to infinity, the strain tends to zero. Therefore, when the stress is removed, the strain in a springpot is always completely recovered; the springpot has \textit{shape memory}, although energy would be dissipated during deformation. The greater the power-law exponent of the springpot $\beta$, the larger the time required for total strain recovery, as shown in figure \ref{fig:f3} (b). For example, by looking at the response of a springpot with $\beta = 0.7$ (green curve in figure \ref{fig:f3} (b)) over a finite time interval, one might qualitatively interpret the slow recovery process as evidence for plastic behaviour -- although this is erroneous because strain eventually converges to zero.
This shows that long experiments are needed to fully capture the rheology of power-law materials.

\begin{figure}[!t]
  \centering
 \includegraphics[width=1\textwidth]{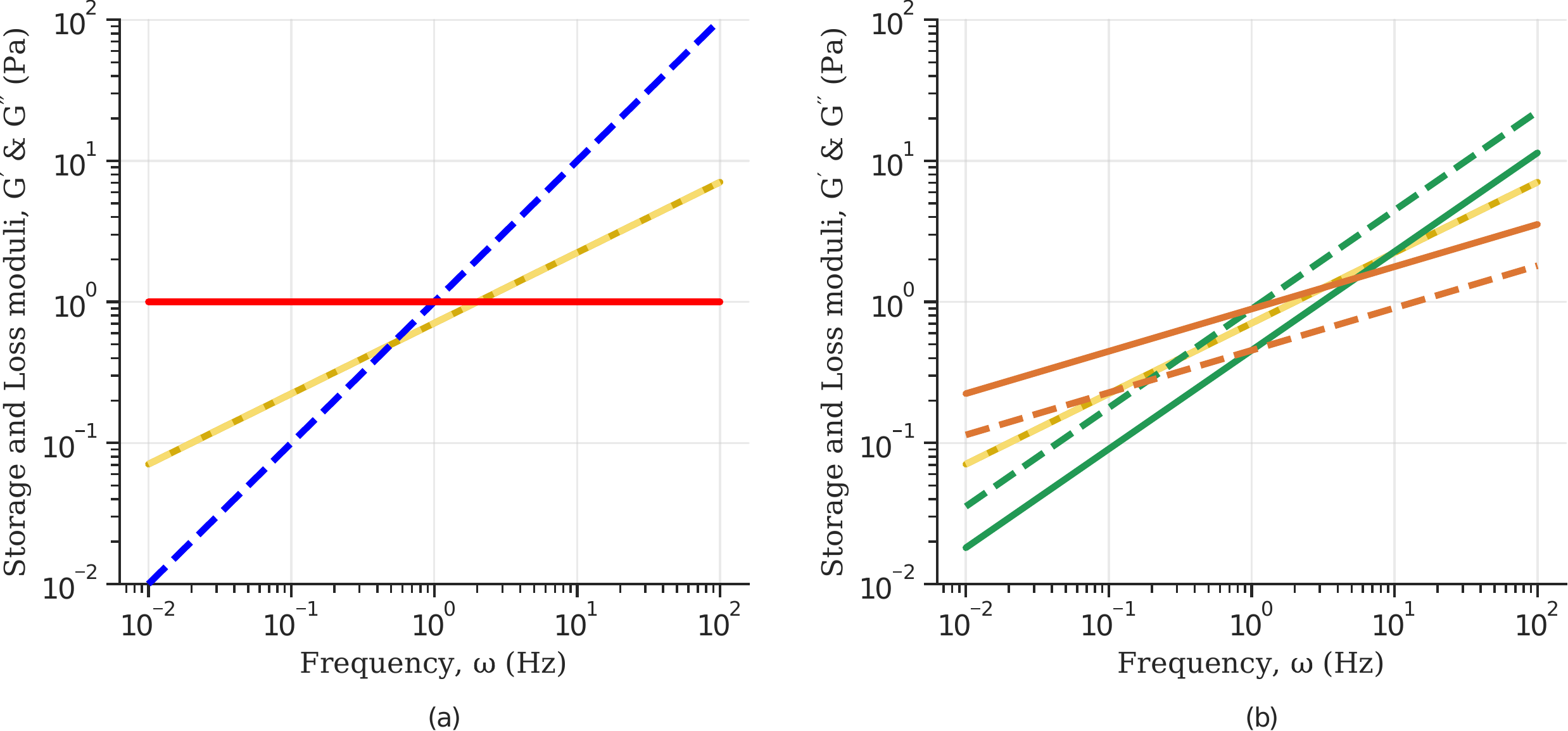}
  \caption{Storage (solid line) and loss (dash line) moduli for: (a) spring (red curve), dashpot (blue curve) and the special case of $\beta = 0.5$ (yellow curves) for which storage and loss moduli are exactly the same; (b) for $\beta< 0.5$ the storage modulus is always greater than the loss modulus ($\beta = 0.3$, orange curves), whilst the opposite is true for $\beta > 0.5$ ($\beta = 0.7$, green curves).}
  \label{fig:f5}
\end{figure}

 It is also illuminating to consider the response of the springpot in the frequency domain. However, the Caputo definition for the fractional derivative is not appropriate here since the assumption that $\epsilon(t)=0$ for $t<0$ does not hold anymore if we excite the springpot with a periodic deformation. To overcome this problem, we use the Generalised Liouville-Caputo formulation 
\begin{equation}\label{eq:lio}
	^{L}D^{\beta}f(t) = \frac{1}{\Gamma(1-\beta)} \int_{-\infty}^{t} (t-\tau)^{-\beta} \frac{df(\tau)}{d\tau} d\tau,
\end{equation}
where $L$ denotes the Generalised Liouville-Caputo fractional derivative. Note that this definition is equivalent to the Caputo Fractional Derivative for functions $f$ such that $f(t<0)=0$ \cite{ortigueira2020fractional}. By now considering the oscillatory load $\epsilon(t) = e^{i \omega t}$ with equation \ref{eq:lio} (where $^{L}D^{\beta}f(t) = \sigma(t)$ and $f(t) = \epsilon(t)$), the complex modulus of a springpot can be found
\begin{equation}\label{eq:19}
	G^{*}(\omega) =  c_\beta (\omega \mathrm{i}\mkern1mu)^{\beta} = c_\beta \omega^{\beta} e^{i\frac{\pi}{2}\beta},
\end{equation}
where $\omega$ is the frequency. Separating the real and imaginary parts, we find the storage and loss moduli respectively
\begin{equation}\label{eq:20}
  \begin{aligned}
	G^{\prime}(\omega) =  \Re(G^{*}) = c_\beta \omega^\beta \cos(\frac{\pi}{2}\beta),\\
    G^{\prime\prime}(\omega) =  \Im(G^{*}) = c_\beta \omega^\beta \sin(\frac{\pi}{2}\beta).
  \end{aligned}
\end{equation}
They follow a power-law behaviour with the same exponent $\beta$. When $\beta = 0$ the  loss modulus is zero as in the case of the spring, whilst when $\beta = 1$ the storage modulus is zero as in the case of a dashpot (figure \ref{fig:f5} (a)). The storage and loss modulus exactly match when $\beta=0.5$ (figure \ref{fig:f5} (a)). For $\beta< 0.5$ the storage modulus is always greater than the loss modulus, whilst the opposite is true for $\beta > 0.5$ (figure \ref{fig:f5} (b)). The phase angle $\delta$ between the excitation and the response is related to the storage and loss moduli by $\tan(\delta) = G^{\prime\prime}/G^{\prime}$, from which we can derive the retardation phase for a springpot as $\delta = \frac{\pi}{2}\beta$, constant for all frequencies.

\subsubsection{Examples of the use of the springpot in practical cases}

The first material investigated using fractional calculus was bitumen in 1944, in a study conducted by Scott-Blair and Veinoglou \cite{blair1944study}. Since then, use of the springpot has been somewhat erratic, possibly because of its non-trivial mathematical foundations. However, the springpot has found use in a diverse array of materials outside of the geological and construction-materials contexts; for example, in gels. An early example is the work of Winter and Chambon \cite{winter1986analysis} who derived a springpot-like modulus for crosslinking polymers at their gelation point, which was used for analysis of polydimethylsiloxane gel data. More recently, a springpot has been used in modelling asphalt mixtures \cite{mino2016linear} and to capture the oscillatory rheological response of agarose, a polysaccharide extracted from red algae \cite{chen2004dynamic}.

More examples of springpot usage can be found in the context of biological materials more generally. One of the first applications of the springpot in tissue biomechanics was for the study of the viscoelastic properties of lung \cite{suki1994lung}. More recently, it has been used to capture the relaxation behaviour of human arteries \cite{craiem2008fractional} and the dynamic (oscillatory) response of brain tissue for the understanding of neurological disorders \cite{nicolas2018biomechanical}. There are also several biomechanical studies which make use of a power-law empirical function that corresponds to one of the three springpot moduli discussed above. For example, the empirical function used to analyze the creep response of single cells can be easily related to the springpot's creep modulus \cite{desprat2005creep,zhou2010power,pullarkat2007rheological}. Similarly, an empirical function used to analyze the relaxation response of smooth muscle cells is equivalent to that of a single springpot \cite{hemmer2009role}.

\subsection{Generalised fractional viscoelastic models}

Many materials exhibit power-law viscoelastic behaviour. However, the power-law regime is often limited in time; some materials show a power-law response at short time-scale that converges to a well-defined plateau (solid like behaviour) \cite{khalilgharibi,craiem2010fractional}, whilst others may exhibit a power-law response followed by a continuous flow of the material (fluid like behaviour) \cite{tripathi2010peristaltic,celauro2012experimental}. In order to capture the diversity of behaviours, \textit{ad-hoc} empirical moduli and large networks of springs and dashpots (i.e. exponential terms) have been often used. However, as previously discussed, these approaches each have disadvantages.

A third useful approach is to combine springpots with themselves and with traditional spring and dashpot elements. In this way, the many advantages of the spring-dashpot network approach can be extended to concisely capture power-law viscoelastic regimes. Such models are often referred to as \textit{generalised viscoelastic models} and they have been mathematically characterized in the past \cite{schiessel1995generalized,mainardi2011creep,pritchard2017oscillations}. The two simplest configurations are two springpots in series or in parallel, which are the fractional analogues of the Maxwell and Kelvin-Voigt models respectively. Examples of the use of the generalized models in practical cases are reported in table \ref{tab:main} (see \cite{lai2016investigation} for a further review of fractional viscoelasticity in geotechnical engineering). In the following section, we demonstrate the behaviour diversity that these generalised models are capable of.

\begin{figure}[!t]
  \centering
 \includegraphics[width=0.9\textwidth]{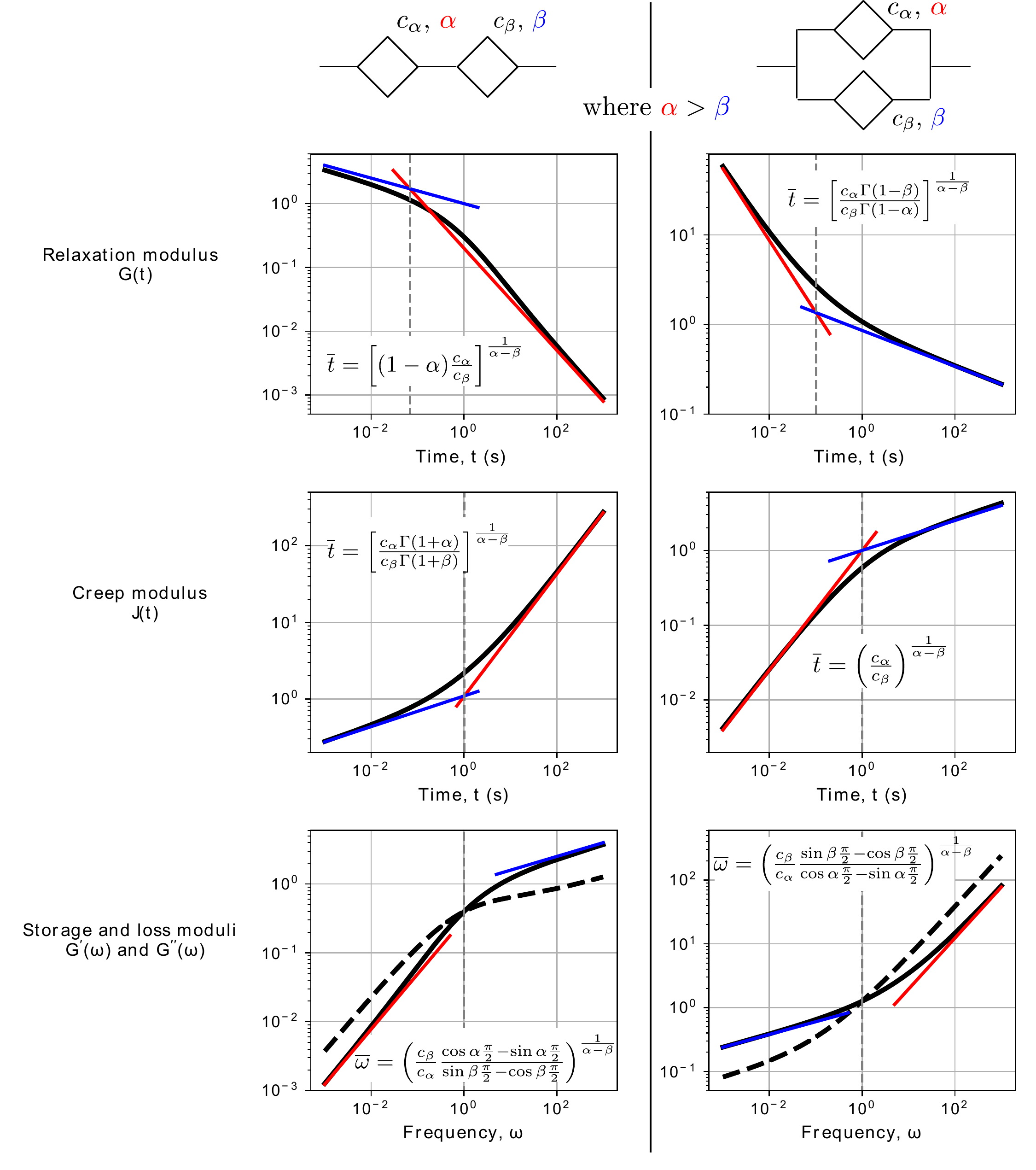}
  \caption{Qualitative behaviour of two springpots in series and parallel. When two springpots are placed in series, the short time scale response is dominated by the springpot with lower exponent, while the long time-scale response is controlled by the springpot with higher exponent (top figures). The contrary is true for two springpots in parallel (bottom figures). The parameters of the models are $c_\alpha$ = 1, $\alpha$ = 0.8, $c_\beta$ = 1, $\beta$ = 0.2.}
  \label{fig:f15}
\end{figure}

\subsubsection{The Fractional Maxwell and Kelvin-Voigt Models: Heuristic Overview}

The choice of which viscoelastic model to use is largely informed by the qualitative behaviour of the experimental data. A key advantage of schematically representing viscoelastic models as networks of elements is that it facilitates visual intuition for a model's behaviour. Despite the mathematical complexity of the springpot, limit behaviours can be obtained as commonly done for springs and dashpots models. To address this, we provide below insight on the qualitative behaviour of the two simplest generalised fractional viscoelastic models, which can be extrapolated to more complex models.

When two springpots of exponents $\alpha$ and $\beta$ (with $\alpha>\beta$) are placed in \textbf{series}, the element with the lower exponent (more ``springy'') dominates the short time-scale response, while the springpot with the higher exponent determines the long time-scale behaviour (see figure \ref{fig:f15} first column). This is true for both the relaxation and creep responses, thus
\begin{equation}
G(t)  \sim \begin{cases}t^{-\beta} & t  \rightarrow  0\\t^{-\alpha} & t  \rightarrow   \infty \end{cases}\quad \textrm{and} \quad J(t)  \sim \begin{cases}t^{\beta} & t  \rightarrow  0\\t^{\alpha} & t  \rightarrow   \infty \end{cases} \quad\textrm{where} \quad 1>\alpha>\beta.
\end{equation}
When two springpots are combined in \textbf{parallel} the opposite is true: both relaxation and creep responses are dominated by the springpot with higher exponent at short time scale, while the springpot with the lower exponent dominates at long time-scale (see figure \ref{fig:f15} second column), thus
\begin{equation}
G(t)  \sim \begin{cases}t^{-\alpha} & t  \rightarrow  0\\t^{-\beta} & t  \rightarrow   \infty \end{cases}\quad \textrm{and} \quad J(t)  \sim \begin{cases}t^{\alpha} & t  \rightarrow  0\\t^{\beta} & t  \rightarrow   \infty \end{cases} \quad\textrm{where} \quad 1>\alpha>\beta.
\end{equation}
Indeed, traditional viscoelastic models can be seen as special cases of the generalised fractional models obtained by specialising the springpots to either springs or dashpots. The qualitative behaviours reported above are consistent with those observed in conventional viscoelastic models, where the Maxwell model shows a liquid-like behaviour at long time scale (dominated by the dashpot, $\alpha$ = 1), while the Kelvin-Voigt model shows a solid-like behaviour at long time-scale (dominated by the spring, $\beta$ = 0). 

Although the above heuristics directly apply to the two-springpot models, the insights can be extrapolated to more complex models in a straightforward manner. For example, consider the Standard Linear Solid model (figure \ref{fig:04}) but with the spring $k_1$ replaced by a springpot. If we look at the global network, the $k_2$ spring in parallel dominates the long-time scale response given its lower power. However, if we focus only on the upper arm, the model will follow the series configuration behaviour. Thus, the short time-scale response is dominated by the springpot and the intermediate response by the dashpot. A more graphical description of this heuristic process has been recently presented by Bonfanti \textit{et al.} \cite{bonfanti2019unified}. To build further intuition, readers may also refer to the Annex, where the relaxation and creep behaviours of more complicated models are briefly reported.

\renewcommand\arraystretch{1.2}

\begin{table}[t]
\caption{Examples of practical uses of generalized models. }
\centering
\begin{tabular*}{1.0\textwidth}{l@{\extracolsep{\fill}}l@{\extracolsep{\fill}}}
\toprule
\textbf{Model} & \multicolumn1{p{120mm}}{\textbf{Applications}}\\
\midrule
\multirow{6}{25mm}{Fractional Kelvin-Voigt} & \multicolumn1{p{120mm}}{Modelling of human prostate tissue to develop novel criteria for cancer detection \cite{zhang2008quantitative}.} \\
 & \multicolumn1{p{120mm}}{Modelling of the oscillatory response of canine liver \cite{kiss2004viscoelastic}.} \\
 & \multicolumn1{p{120mm}}{Modelling of the relaxation response of beast cells and tissue samples \cite{zhang2018modeling}.} \\
& \multicolumn1{p{120mm}}{Modelling of the tissue mimicking materials CF-11 and gelatin \cite{meral2010fractional}.} \\
& \multicolumn1{p{120mm}}{Modelling of the viscoelastic response of potato starch gel \cite{choudhury2012forced}.} \\
& \multicolumn1{p{120mm}}{Modelling of a three-dimensional particulate gel \cite{bouzid2018computing}.} \\
& \multicolumn1{p{120mm}}{Studying of the applicability of the model to wellbore creep \cite{yu2017wellbore} } \\
\midrule

\multirow{5}{25mm}{Fractional Maxwell}  & \multicolumn1{p{120mm}}{Modelling of the creep behaviour of rocks \cite{wu2015improved}.} \\
& \multicolumn1{p{120mm}}{Modelling of the viscoelastic response of tight sandstone \cite{ding2017unexpected}.} \\
&  \multicolumn1{p{120mm}}{Modelling of both colloidal \cite{aime2018power} and carbopol \cite{lidon2017power} gel rheology.} \\
&  \multicolumn1{p{120mm}}{Modelling of arterial tissue \cite{shen2013fractional}.} \\
&  \multicolumn1{p{120mm}}{Modelling the properties of food gels \cite{faber2017describing}.} \\
&  \multicolumn1{p{120mm}}{Modelling the mechanical properties of collagen gel \cite{holder2018control}.} \\
\midrule

\multirow{7}{25mm}{Fractional Standard Linear Solid model} & \multicolumn1{p{120mm}}{Study of single red blood cells \cite{craiem2010fractional}.}\\
& \multicolumn1{p{120mm}}{Modelling of breast cancer cells to develop new diagnostic tool \cite{carmichael2015fractional}.} \\
& \multicolumn1{p{120mm}}{Modelling of artery walls for the study of aneurysms \cite{perdikaris2014fractional,yu2016fractional}.} \\
& \multicolumn1{p{120mm}}{Quantification of changes in viscoelastic response of lung parenchyma due to trauma \cite{dai2015model}.} \\
& \multicolumn1{p{120mm}}{Modelling of the oscillatory response of cancerous cells; the information was then used to selectively attack malignant cells during treatments \cite{fraldi2015frequency}.} \\
& \multicolumn1{p{120mm}}{Modelling of the viscoelastic response of brain tissue \cite{kohandel2005frequency}.}\\
& \multicolumn1{p{120mm}}{Modelling of the viscoelastic properties of the glassy amorphous polymer poly-methyl-methacrylate \cite{alcoutlabi1998application}.} \\
\midrule

\multirow{1}{25mm}{Fractional Burgers}  &  \multicolumn1{p{120mm}}{Viscoelastic analysis of waxy crude oil \cite{zhang2016rheological,hou2014new}.} \\[2.5ex]

\bottomrule
\end{tabular*}
\label{tab:main}
\end{table}

\begin{figure}[t]
  \centering
 \includegraphics[width=0.9\textwidth]{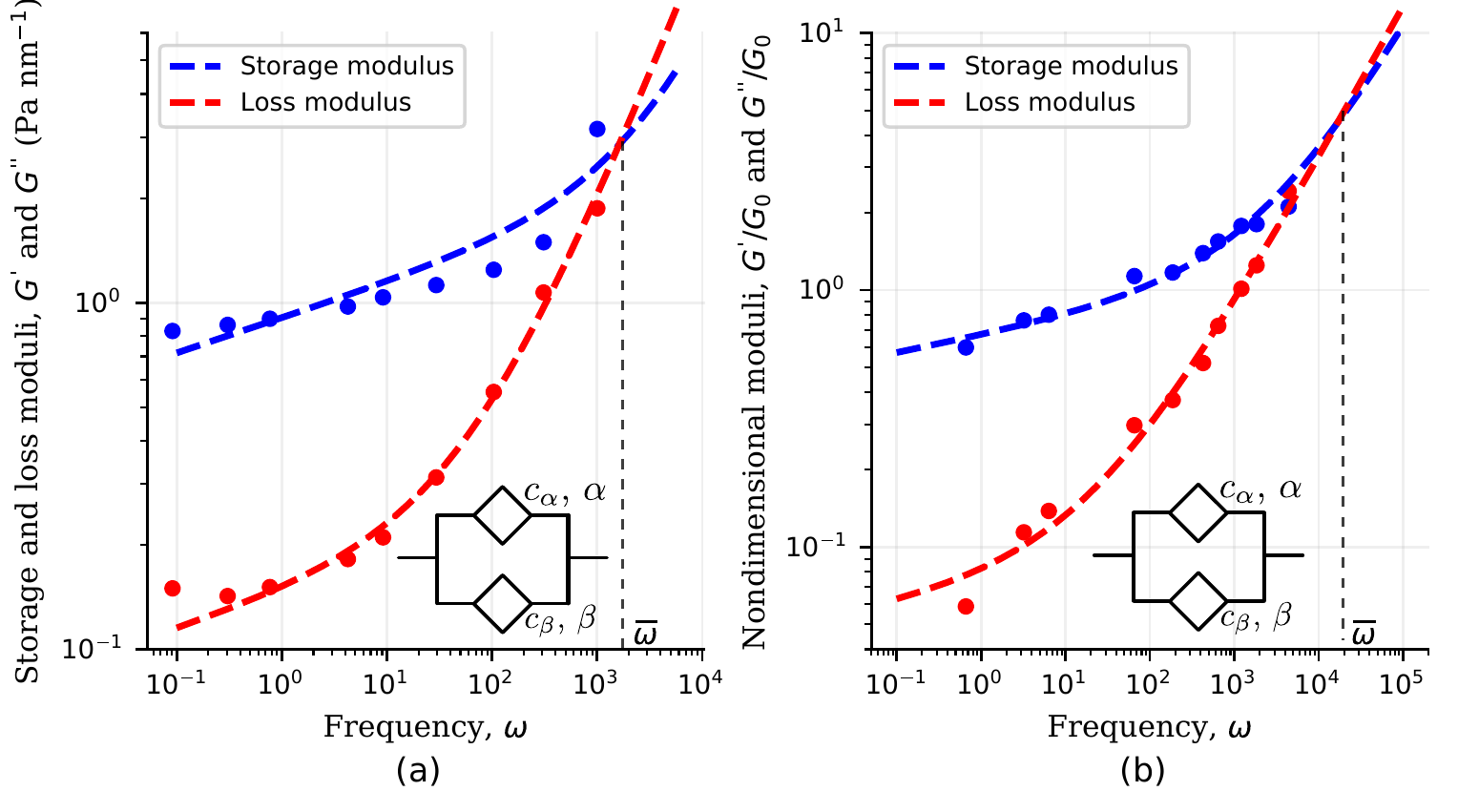}
  \caption{Fitting of storage modulus and loss modulus showing two power-law regimes using Fractional Kelvin-Voigt model. Dynamic response of (a) smooth muscle cells cytoskeleton \cite{deng2006fast} and (b) kidney epithelial cells ATP-depleted \cite{hoffman2006consensus}. Fitted model parameter values in table \ref{tab:1}.}
  \label{fig:f8}
\end{figure}

\vspace*{0.5cm}

In the following sections we demonstrate the benefits of using fractional models in a wide range of practical applications. 
We first show that a number of empirical functions previously introduced in the literature actually possess an equivalent fractional model.
We then demonstrate how the use of fractional models often allows us to capture the mechanical behaviour of power-law materials more accurately than spring-dashpot viscoelastic models while using less parameters.

\subsubsection{Examples of empirical functions equivalent to fractional viscoelastic models}

\paragraph{Structural damping model.} A number of dynamic (oscillatory) viscoelastic tests on biological tissue were fitted with an empirically derived model known as the \textit{structural damping} (or hysteretic damping) model \cite{fredberg1989imperfect,rother2015cytoskeleton,cai2013quantifying,fabry2001scaling,fabry2003time,smith2005probing,cai2017temporal} whose complex modulus given by:
\begin{equation}\label{eq:21}
	G^{*}(\omega) =  \mu \left(\mathrm{i}\omega \right)+G_0 \left(\frac{\omega}{\omega_0}\right)^\beta \left(1+\overline{\eta}\mathrm{i}\right) \cos\left(\beta \frac{\pi}{2}\right)  ,
\end{equation}
where $\overline{\eta} = \tan\left(\beta \frac{\pi}{2}\right)$ is commonly called the hysteresivity or structural damping coefficient, $\beta$ is the power-law exponent, $G_0$ and $\omega_0$ are two scaling factors for stiffness and frequency respectively, and $\mu$ the Newtonian viscosity.
This modulus can be shown to be exactly equivalent to that of a semi-fractional Kelvin-Voigt model consisting of a springpot in parallel with a dashpot. In fact, equation \eqref{eq:21} can be re-written as:
\begin{equation}\label{eq:22}
	G^{*}(\omega) =  \mu \left(\mathrm{i}\omega \right)+\frac{G_0}{\omega_0^\beta}  \left(\mathrm{i}\omega\right)^{\beta},
\end{equation}
which is equivalent to the complex modulus of a dashpot in parallel to a springpot $G^{*}(\omega) =  c_\alpha \left(\mathrm{i}\omega\right)^{\alpha} + c_\beta \left(\mathrm{i}\omega\right)^{\beta}$, when $\alpha=1$, $c_\alpha = \mu$ and $c_\beta = G_0/ \omega_0^\beta$.

\paragraph{Two power-law regime.} The frequency response of single cells often exhibits two power-law regimes \cite{hoffman2006consensus,deng2006fast}. A lower-power exponent at low frequencies, followed by a higher power-law exponent at high frequencies. Such behaviour has been frequently been analyzed by the empirically derived superposition of two power-laws
\begin{equation}\label{eq:23}
	G^{*}(\omega) =  A (\mathrm{i}\omega)^\alpha + B (\mathrm{i}\omega)^\beta,
\end{equation}
where one exponent is often fixed to $\beta = 3/4$. The expression above is exactly equivalent to the complex modulus of the Fractional Kelvin-Voigt model shown in the inset of figure \ref{fig:f8} and consisting of two springpots in parallel. Examples of fitting the storage and loss moduli of single cells to the Fractional Kelvin-Voigt model are shown in figure \ref{fig:f8}.

\begin{figure}[t]
  \centering
 \includegraphics[width=0.9\textwidth]{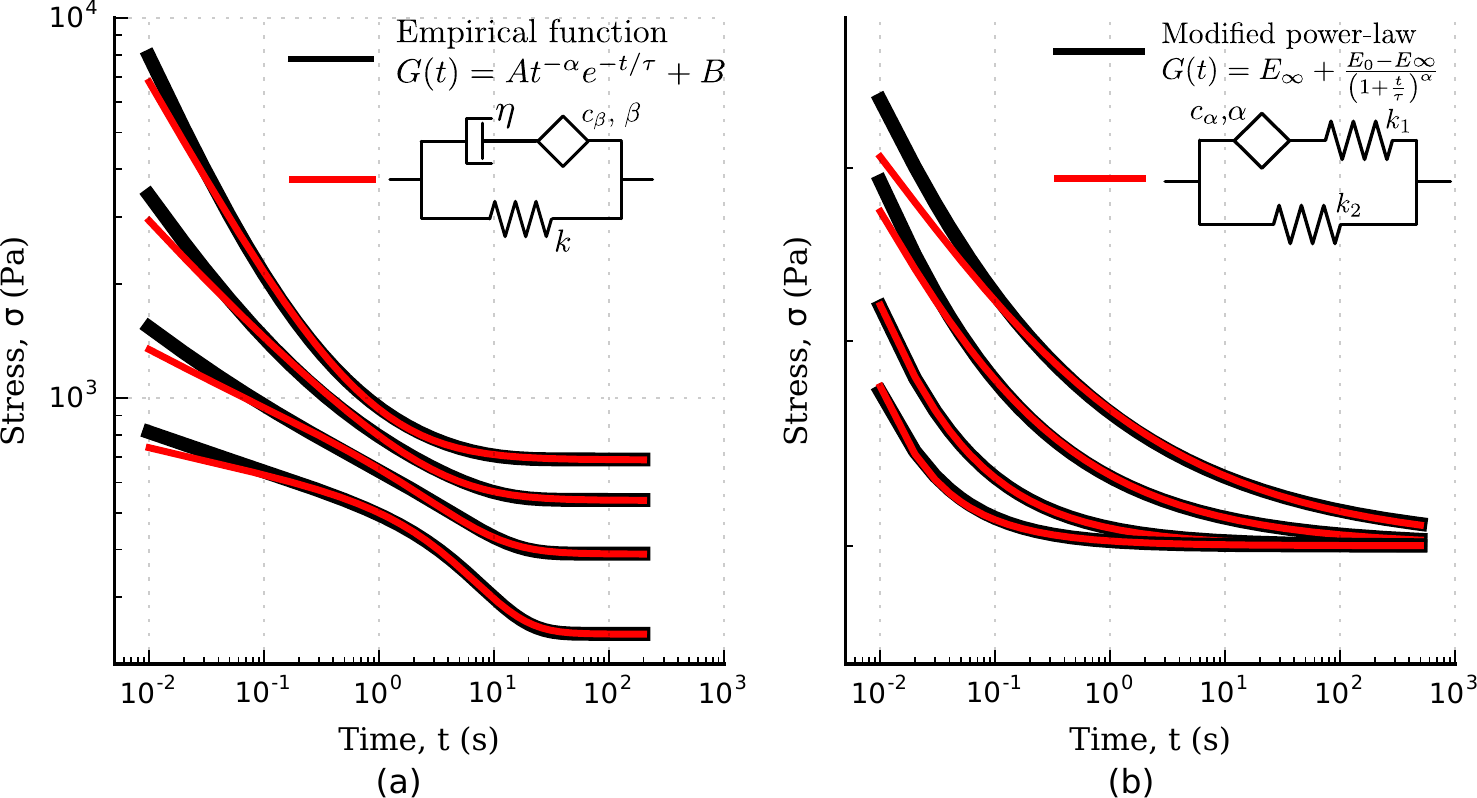}
  \caption{Fitting of empirical functions with fractional models. (a) Biphasic empirical function describing the relaxation response of epithelial monolayers fitted with the solid fractional model for different model parameters (fitted models parameters in table \ref{tab:2}). (b) Modified power-law empirical model for the relaxation response fitted with the Fractional Standard linear solid model (fitted model parameters in table \ref{tab:5}).}
  \label{fig:f9}
\end{figure}

\begin{figure}[t]
  \centering
 \includegraphics[width=1\textwidth]{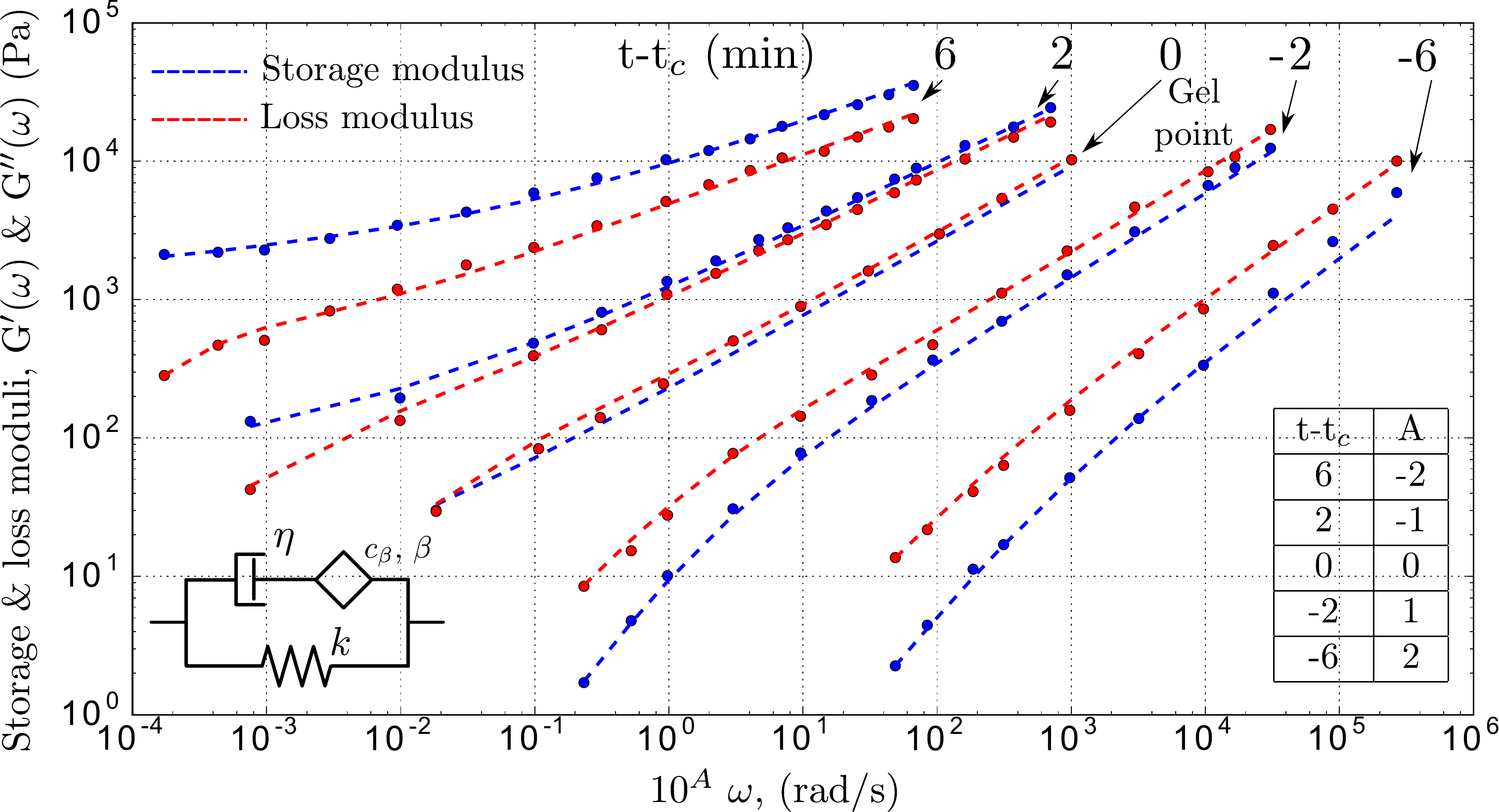}
  \caption{Storage (blue) and loss (red) moduli of cross-linked PDMS samples at different extent of reactions ($\rm{t}-\rm{t_c}$, where t is the current time and $\rm{t_c}$ the time of gelation). The \textit{x}-axis is shifted by A decades to facilitated reading. The dots are the experimental data (from \cite{friedrich1988extension}), while the dashed lines the fitted fractional viscoelastic model shown in the inset. Fitted parameters reported in table \ref{tab:4} in Appendix. }
  \label{fig:f16}
\end{figure}

\paragraph{Power-law cut-off.} Recently, Khalilgharibi et.al. \cite{khalilgharibi} studied the relaxation response of epithelial monolayers, a sheet of cells devoid of substrate, to uncover the subcellular components responsible for stress dissipation. The relaxation response consists of an initial power-law phase in the first 5 s, followed by an exponential phase that reaches a plateau at $\sim60$ s. This biphasic relaxation response was captured by extending the power-law empirical function to include an exponential regime
\begin{equation}\label{eq:emp}
	G(t) =  A t^{-\alpha} {\rm e}^{-t / \tau} + B.
\end{equation}
More recently however, traditional viscoelastic elements have been combined with a springpot to characterize the viscoelastic response of epithelial monolayers at both short and long time-scale \cite{bonfanti2019unified}. The novel configuration features a springpot that sustains all the load at short time-scale, after which the load is then slowly transferred to a dashpot placed in series which gives rise to the exponential behaviour. This dissipative branch is placed in parallel with a spring to obtain the final plateau (inset in figure \ref{fig:f9} (a)). The mathematical expression for the relaxation modulus of the fractional model introduced by Bonfanti \textit{et al.} \cite{bonfanti2019unified} is
\begin{equation}\label{eq:11}
	G(t) =  c_\beta t^{-\beta} E_{1-\beta,1-\beta}\left(-\frac{c_\beta}{\eta} t^{1-\beta}\right) + k,
\end{equation}
where $c_\beta$ and $\beta$ are the springpot parameters, $\eta$ the dashpot viscosity, $k$ the spring stiffness, and $E_{a,b}\left(z\right)$ is the Mittag-Leffler function, a special function that often arises from the solution of fractional differential equations \cite{haubold2011mittag,gorenflo2002computation}. By fitting the fractional model to the empirical functions we can demonstrate that they are approximately equivalent (figure \ref{fig:f9} (a)) and therefore the empirical function can be mapped into the fractional viscoelastic framework.

Interestingly, the same empirical model was independently developed to capture the relaxation response of one of the most widely used elastomers, polydimethylsiloxane (PDMS) \cite{friedrich1988extension}, and the rheological behaviour of alginate-based gels \cite{nobile2008development,moresi2007characterisation}. The study of the liquid-solid transition during gelation has attracted significant attention in the past \cite{winter1997rheology}. Friedrich \text{et al.} \cite{friedrich1988extension} reported the rheological response of stoichiometrically balanced polydimethylsiloxane (PDMS) in which the cross-linking reaction was interrupted at different times, before and after the critical point (referred to as the gel point). As shown in figure \ref{fig:f16}, the four-parameter fractional solid model can also be applied here to accurately captures the behaviour of the time-evolving cross-linking reaction pre- and post-gelation. 

Physically, during gelation a polymer network forms. The critical gel point is associated with an abrupt change in viscosity \cite{odian2004principles}, which is defined as the creation of the first percolation cluster that spans the sample. By observing the values of the fitted parameters reported in table \ref{tab:4} in Appendix, the fluid-like behaviour of the PDMS pre-gel is confirmed by the zero value of the stiffness of the spring in parallel. As expected from physical considerations, at the gel point we observe a rapid increase of the viscosity (figure \ref{fig:params} in Appendix). Post-gel the PDMS behaves as a solid, which is confirmed by the rapid increase of the stiffness $k$. As discussed above, the storage and loss moduli of the springpot are exactly equal when $\beta$ = 0.5, or physically, when the springpot is exactly intermediate between a liquid and a solid. The physical relevance of this parameter is confirmed from the fitted springpot coefficient shown in table \ref{tab:4}, where $\beta$ at time $t - t_c = 0$ (gel point) is $\sim 0.5$.

\paragraph{Modified power-law models.} Another empirical model, recently used for the analysis of the relaxation response of various benign and malignant cell lines, is the modified power-law (MPL) model \cite{efremov2017measuring} which can be written as
\begin{equation}\label{eq:MPL}
    E(t) = E_{\infty } + \frac{E_{0} - E_{\infty}}{(1 + \frac{t}{\tau})^{\alpha}},
\end{equation}
where $E_{0}$ is the instantaneous or `glassy' modulus, $E_{\infty}$ is the plateau or `rubbery' modulus, $\tau$ is a time scaling, and $\alpha$ determines the power-law gradient of relaxation. In contrast to a regular power-law model, the MPL model has the convenient property of being well defined at short time scales. Its behaviour is notably similar to a Fractional Standard Linear Solid model (which is equivalent to a Fractional Zener model with two springpots specialised to springs), as shown in figure \ref{fig:f9} (b). The relaxation modulus of the Fractional Standard Linear Solid model is defined as
\begin{equation}
    G(t) = k_{\beta} E_{\alpha',1} \left( -\frac{k_{\beta}}{c_{\alpha'}}t^{\alpha'} \right) + k_{\gamma}
\end{equation}
where $k_{\beta}$ and  $k_{\gamma}$ are the two spring constants, $c_{\alpha'}$ and $\alpha'$ are the springpot parameters (with a prime added to $\alpha$ to avoid confusion with the MPL $\alpha$ parameter), and $E$ is the Mittag-Leffler function \cite{haubold2011mittag,gorenflo2002computation}. The similarity between the two models was also discussed by Bagley \cite{bagley1989power}, who made comparisons between their relaxation spectra. Here we show their similarity by directly matching their boundary condition parameters at $t=0$ and $t \to \infty$ and fitting the Fractional Standard Linear Solid parameters $c_{\alpha'}$ and $\alpha'$ to the MPL (figure \ref{fig:f9} (b)). The fractional model fits the MPL well, especially at longer time scales where both the Mittag-Leffler function and the MPL asymptotically approach a simple power-law \cite{mainardi2014propMittLeff}. Interestingly, the MPL model has also been used in several asphalt and asphalt-concrete studies \cite{kim1995correspondence,lee1998viscoelastic,forough2016comparing} indicating that the fractional standard linear solid may have utility in this field.

\begin{figure}[t]
  \centering
 \includegraphics[width=0.9\textwidth]{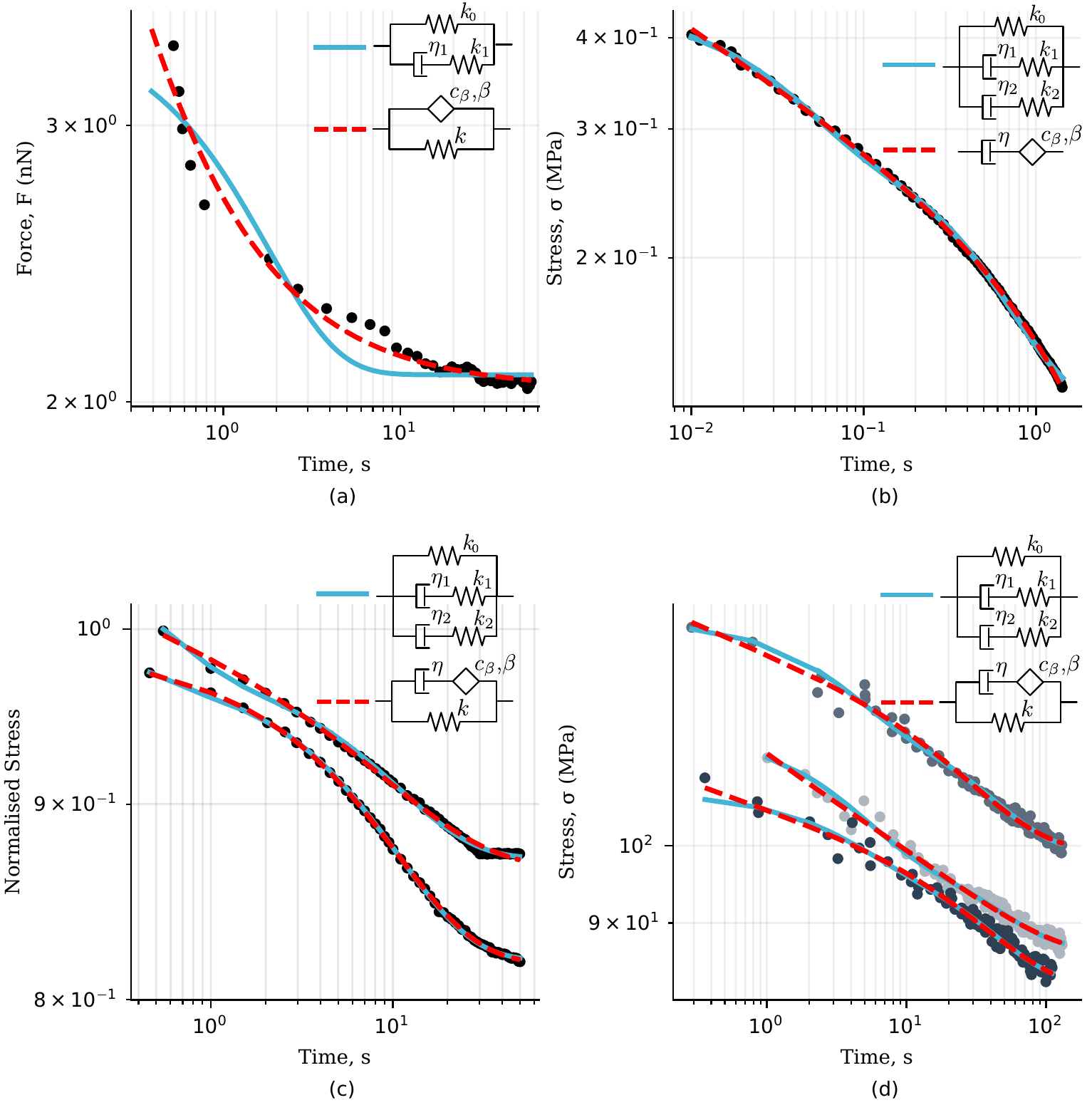}
  \caption{(a) Relaxation data from zonal articular chondrocytes \cite{darling2006viscoelastic} fitted with a standard linear solid and fractional Kelvin-Voigt specialized by one spring. (b) Relaxation data from tomato mesocarp cells \cite{li2016viscoelastic} fitted with a 2 time-scale generalised Maxwell model and a fractional Maxwell model specialized by one dash-pot. (c) Relaxation data from PCL/bio-active glass \cite{shahin2018mechanical} fitted with a 2 time-scale generalised Maxwell model and a fractional special model. (d) Relaxation data from collagen fibrils \cite{shen2011viscoelastic} fitted with a 2 time-scale generalised Maxwell model and a fractional special model. All fitted parameters are reported in table \ref{tab:3}.}
  \label{fig:f10b}
\end{figure}

\subsubsection{Fractional models capture complex power-law behaviours with less parameters}

As discussed, spring-dashpot viscoelastic models can be used to capture power-law viscoelasticity but this results in an abundance of model parameters. We have shown that the springpot provides an efficient way to describe the rheology of power-law materials with only two parameters. To further illustrate the ability of fractional models to accurately capture complex power-law responses with a minimum number of model parameters, we now re-analyse the time-response of different power-law materials, originally modelled using spring-dashpot networks, by using fractional viscoelastic models.

We first analyse the rheological behaviour of single cells presented by Darling \textit{et al.} \cite{darling2006viscoelastic}. They investigated the relaxation response of zonal articular chondrocytes by the use of AFM (Atomic Force Microscopy). The Standard Linear Solid model originally used in the paper struggles to capture the short-time scale response (see figure \ref{fig:f10b} (a)). We propose the use of the fractional Kelvin-Voigt model (springpot in parallel to a spring) that involves the same number of parameters (see inset in figure \ref{fig:f10b} (a)). A cursory look at the comparison between the fits of traditional and fractional viscoelastic models to the relaxation response of zonal articular chondrocytes shown in figure \ref{fig:f10b} (a) reveals the ability of the fractional model to significantly improve the quality of the fitting by accurately capturing the fast relaxation response while using the same number of parameters as the spring-dashpot model used in the original paper. 

Recently, Zhang \text{et al.} \cite{zhang2016rheological} tested tomato mesocarp cells under high speed microcompression to understand how industrial handling may affect the integrity of fresh fruits. To successfully capture the distribution of time-scales that gives rise to the power-law regime at short time-scale, a 2 time-scale Standard Linear Solid that requires five parameters was originally used. By using a fractional Maxwell model (springpot in series to a dashpot) we achieve the same excellent fit to the data whilst reducing the number of parameters to three (figure \ref{fig:f10b} (b)).

The two-time-scale Standard Linear Solid model has recently been applied to the analysis of the rheological behaviour of other materials, such as polycaprolactone bioactive glass tested under compression \cite{shahin2018mechanical} and single collagen fibrils from the extracellular matrix under tensile testing \cite{shen2011viscoelastic}. Given that they present the same qualitative behaviour as the epithelial monolayers (power-law followed by exponential behaviour until a final steady-state value is reached), we have successfully applied the fractional model recently developed by Bonfanti \textit{et al.} \cite{bonfanti2019unified} to data from these other materials. The model accurately fits both short and long time-scale responses while using one less parameters (figure \ref{fig:f10b} (c)-(d)).

\section{Conclusions}

This review has demonstrated the ability of fractional viscoelastic models to accurately capture the rheological responses of a broad range of materials, identifying materials parameters that account for the wide distribution of time-scales involved in power-law behaviour. Despite the limitation to the linear response, fractional models exhibit rich behaviours consistent with empirical data in both relaxation and creep experiment. Looking at large deformations and failure of complex materials remains however a challenge that cannot be tackled by linear rheological models and we anticipate significant efforts to tackle these questions in future.

The main impediments to the use and dissemination of fractional viscoelastic models appears to be their relatively intricate mathematical formalism and the difficulty of their numerical implementation. 
This document illustrated the qualitative response of the springpot element and its composition into simple networks in series and in parallel configurations, helping the reader to build intuition. An annex to the review [REF HERE] provides an exhaustive list of rheological models including up to three elements, alongside the analytical expression of their moduli and graphs illustrating their behaviour. We also developed a software library to facilitate the fitting of experimental data and prediction of power-law behaviours \cite{kaplanrheos}. The software package allows non-experts to fit their own data using a wide selection of viscoelastic models, accounting for complex loading patterns (all figures of this work have been created using RHEOS). Altogether, these elements significantly lower the barriers to using fractional models to analyse the rheology of power-law materials.

Although fractional models extend the range of behaviours that can be modelled with rheological elements, they do not provide on their own  explanations for the underlying mechanisms giving rise to the observed macroscopic power-law behaviour. A deeper understanding of soft material mechanics would require a more systematic theoretical analysis of experimental data that would provide a physical underpinning for the emergence of power-law behaviours. However, fractional models provide systematic approaches to capture material parameters that can be compared across studies, and such comparison is likely to provide a better handle on the underlying physical significance of power-law behaviours.

\section*{Acknowledgment}

The authors wish to acknowledge present and past members of the Kabla and Charras labs for stimulating discussions. AB, JLK, and AJK acknowledge the BBSRC grants BB/M002578/1,BB/K018175/1, and BB/P003184/1. JLK would like thank the George and Lillian Schiff Foundation for the PhD funding which facilitated this project. G.C. was supported by a consolidator grant from the European Research Council (MolCellTissMech, agreement 647186).

\break
\section*{Appendix: Fitted parameters}

\renewcommand\arraystretch{1.2}

\begin{table}[h!]
\caption{Fitted parameters generalised Maxwell models from figure \ref{fig:05} (b)}
\centering
\begin{tabular*}{0.8\textwidth}{l@{\extracolsep{\fill}}ccc}
\toprule
\textbf{Parameters} & \textbf{1 time-scales} & \textbf{2 time-scales} & \textbf{4 time-scales} \\
\midrule
$\boldsymbol{k_1}$ & 0.95  & 0.72  &  0.44\\
\midrule
$\boldsymbol{\eta_1}$  & 13.22  & 17.42 & 28.02\\
\midrule
$\boldsymbol{k_2}$ &  - & 10.29  &15.27  \\
\midrule
$\boldsymbol{\eta_2}$  & - & 12.92 & 2.60\\
\midrule
$\boldsymbol{k_3}$ & -  & -  & 1.61 \\
\midrule
$\boldsymbol{\eta_3}$  & - & - & 22.07\\
\midrule
$\boldsymbol{k_4}$ &  - & -  & 4.98 \\
\midrule
$\boldsymbol{\eta_4}$  & - & - & 7.67\\
\bottomrule
\end{tabular*}
\label{tab:0}
\end{table}

\begin{table}[h!]
\caption{Fitted parameters fractional Kelvin-Voigt model from figure \ref{fig:f8}}
\centering
\begin{tabular*}{\textwidth}{l@{\extracolsep{\fill}}cccc}
\toprule
\textbf{Data} & $\boldsymbol{c}_{\boldsymbol{\alpha}}$& $\boldsymbol{\alpha}$ & $\boldsymbol{c}_{\boldsymbol{\beta}}$  & $\boldsymbol{\beta}$ \\
\midrule
Bovine trachea smooth muscle cells \cite{deng2006fast} &  0.01 $\left(\frac{\rm Pa}{\rm nm} {\rm s}^\alpha\right)$ & 0.78  & 0.9 $\left(\frac{\rm Pa}{\rm nm} {\rm s}^\beta\right)$  & 0.1  \\
\midrule
TC7 kidney epithelial cells ATP-depleted \cite{hoffman2006consensus} & 0.02   & 0.6  & 0.7   & 0.07  \\
\bottomrule
\end{tabular*}
\label{tab:1}
\end{table}

\begin{table}[h!]
\centering
\caption{Parameters of empirical function and three-element fractional model from figure \ref{fig:f9} (a)}
\begin{tabular*}{0.8\textwidth}{l@{\extracolsep{\fill}}cccc}
\toprule
\multicolumn{1}{l}{}     & \multicolumn{2}{c}{\textbf{Empirical model}} & \multicolumn{2}{c}{\textbf{Fractional model}} \\
\midrule
\multirow{3}*{Curve 1}   & $\boldsymbol{A}$            &  960         & $\boldsymbol{c_\beta}$                &    1125           \\
                         & $\boldsymbol{\alpha}$        &  0.15       & $\boldsymbol{\beta}$                 &   0.1             \\
(Bottom)                         & $\boldsymbol{\tau}$         &   8.0      & $\boldsymbol{\eta}$                &     6650          \\
                         & $\boldsymbol{B}$            &   800       & $\boldsymbol{k}$                  &      800           \\
\midrule
\multirow{4}*{Curve 2}   & $\boldsymbol{A}$            &  960         & $\boldsymbol{c_\beta}$                &  1400               \\
                         & $\boldsymbol{\alpha}$        &  0.3         & $\boldsymbol{\beta}$                 &   0.21              \\
                         & $\boldsymbol{\tau}$         & 8.0          & $\boldsymbol{\eta}$                &  6100               \\
                         & $\boldsymbol{B}$            & 1300          & $\boldsymbol{k}$                  &   1300             \\
\midrule
\multirow{4}*{Curve 3}   & $\boldsymbol{A}$            & 960       & $\boldsymbol{c_\beta}$                &  6000               \\
                         & $\boldsymbol{\alpha}$        &  0.5         & $\boldsymbol{\beta}$               & 0.4                \\
                         & $\boldsymbol{\tau}$         &  8.0         & $\boldsymbol{\eta}$                &  2080               \\
                         & $\boldsymbol{B}$            & 1800          & $\boldsymbol{k}$                  &  1800              \\
\midrule
\multirow{3}*{Curve 4}   & $\boldsymbol{A}$            & 960          & $\boldsymbol{c_\beta}$                &  6700               \\
                         & $\boldsymbol{\alpha}$        &  0.7         & $\boldsymbol{\beta}$                 & 0.54                \\
 (Top)                        & $\boldsymbol{\tau}$         & 8.0          & $\boldsymbol{\eta}$                &  3040              \\
                         & $\boldsymbol{B}$            & 2300          & $\boldsymbol{k}$                  & 2300               \\
\bottomrule
\end{tabular*}
\label{tab:2}
\end{table}

\begin{table}[h!]
\centering
\caption{Parameters of empirical function and three-element fractional model from figure \ref{fig:f9} (b)}
\begin{tabular*}{0.8\textwidth}{l@{\extracolsep{\fill}}cccc}
\toprule
\multicolumn{1}{l}{}     & \multicolumn{2}{c}{\textbf{Empirical model}} & \multicolumn{2}{c}{\textbf{Fractional model}} \\
\midrule
\multirow{3}*{Curve 1}   & $\boldsymbol{E_\infty}$            &  0.5         & $\boldsymbol{c_\alpha}$                &   0.007           \\
                         & $\boldsymbol{E_0}$        &  1.0                 & $\boldsymbol{\alpha}$                 &   0.72             \\
(Bottom)                 & $\boldsymbol{\tau}$         &   1e-3               & $\boldsymbol{k_1}$                &     0.5          \\
                         & $\boldsymbol{\alpha}$            &   0.8          & $\boldsymbol{k_2}$                  &    0.5           \\
\midrule
\multirow{4}*{Curve 2}   & $\boldsymbol{E_\infty}$            &  0.5       & $\boldsymbol{c_\alpha}$                &  0.017               \\
                         & $\boldsymbol{E_0}$        &  1.0                  & $\boldsymbol{\alpha}$                 &   0.59              \\
                         & $\boldsymbol{\tau}$         & 1e-3                 & $\boldsymbol{k_1}$                &  0.5               \\
                         & $\boldsymbol{\alpha}$            & 0.6           & $\boldsymbol{k_2}$                  &  0.5             \\
\midrule
\multirow{4}*{Curve 3}   & $\boldsymbol{E_\infty}$            & 0.5       & $\boldsymbol{c_\alpha}$                &  0.051               \\
                         & $\boldsymbol{E_0}$        &  1.0               & $\boldsymbol{\alpha}$               & 0.42                \\
                         & $\boldsymbol{\tau}$         &  1e-3             & $\boldsymbol{k_1}$                &  0.5               \\
                         & $\boldsymbol{\alpha}$            & 0.4        & $\boldsymbol{k_2}$                  &  0.5              \\
\midrule
\multirow{3}*{Curve 4}   & $\boldsymbol{E_\infty}$            & 0.5          & $\boldsymbol{c_\alpha}$                &  0.096               \\
                         & $\boldsymbol{E_0}$        &  1.0                  & $\boldsymbol{\alpha}$                 & 0.33                \\
 (Top)                   & $\boldsymbol{\tau}$         & 1e-3                 & $\boldsymbol{k_1}$                &   0.5              \\
                         & $\boldsymbol{\alpha}$            & 0.3           & $\boldsymbol{k_2}$                  & 0.5              \\
\bottomrule
\end{tabular*}
\label{tab:5}
\end{table}

\begin{table}[h!]
\centering
\caption{Fitted parameters of the traditional viscoelastic models and the fractional model from figure \ref{fig:f10b}. }
\begin{tabular*}{0.8\textwidth}{l@{\extracolsep{\fill}}llll}
\toprule
\multicolumn{1}{l}{\textbf{Material}}     & \multicolumn{2}{l}{\textbf{Traditional model}} & \multicolumn{2}{l}{\textbf{Fractional model}} \\
\midrule
\multirow{3}*{Zonal articular } & $\boldsymbol{k_0}$ (nN) &260 & $\boldsymbol{c_\beta}$ (nN s$^{\beta}$) & 420 \\
           & $\boldsymbol{\eta_1}$ (nN s)    &  172       & $\boldsymbol{\beta}$     &   0.82       \\
 chondrocytes \cite{darling2006viscoelastic}  & $\boldsymbol{k_1}$  (nN)&   260   & $\boldsymbol{k}$  (nN) &     253     \\
\toprule
\multirow{5}*{Tomato mesocarp }   & $\boldsymbol{k_0}$ (Pa)&   8e5 & $\boldsymbol{\eta}$ (Pa s) &   6e6 \\
                         & $\boldsymbol{\eta_1}$ (Pa s)   &   1e6        & $\boldsymbol{c_\beta}$ (Pa s$^{\beta}$)           &   1.5e6        \\
                         & $\boldsymbol{k_1}$ (Pa)  &  6e5         & $\boldsymbol{\beta}$          &     0.16              \\
cells \cite{li2016viscoelastic}  & $\boldsymbol{\eta_2}$ (Pa s)   &  1e6         &                 &                  \\
                         & $\boldsymbol{k_2}$ (Pa)        &   4e4        &                 &                  \\
\toprule
\multirow{3}*{PCL/bio-active}   & $\boldsymbol{k_0}$    &  0.8  & $\boldsymbol{\eta}$   &  1.6       \\
                         & $\boldsymbol{k_1}$         & 0.1          & $\boldsymbol{c_\beta}$    &   0.15                \\
glass \cite{shahin2018mechanical} & $\boldsymbol{\eta_1}$       &  0.01         & $\boldsymbol{\beta}$              &  0.02                 \\
Bottom sample            & $\boldsymbol{k_2}$         &   0.15        & $\boldsymbol{k}$                  &  0.76                \\
                         & $\boldsymbol{\eta_2}$         &   1.46        &                 &                  \\
\midrule
\multirow{3}*{PCL/bio-active}   & $\boldsymbol{k_0}$    &  0.8         & $\boldsymbol{\eta}$   &  1.5         \\
                         & $\boldsymbol{k_1}$         &   0.1        & $\boldsymbol{c_\beta}$                  &  0.13                 \\
glass \cite{shahin2018mechanical}                         & $\boldsymbol{\eta_1}$          & 0.04          & $\boldsymbol{\beta}$              &  0.13                 \\
Top sample              & $\boldsymbol{k_2}$             &   0.1        & $\boldsymbol{k}$                  &   0.8               \\
                         & $\boldsymbol{\eta_2}$             &  1.02         &                 &                  \\
\toprule
\multirow{4}*{Collagen fibrils \cite{shen2011viscoelastic}}   & $\boldsymbol{k_0}$ (MPa) &  473         & $\boldsymbol{\eta}$ (MPa s)&  4700                 \\
                         & $\boldsymbol{k_1}$ (MPa)        &   93        & $\boldsymbol{c_\beta}$  (MPa s$^{\beta}$)                &  165                 \\
                         & $\boldsymbol{\eta_1}$  (MPa s)       & 3800          & $\boldsymbol{\beta}$              &  0.12                \\
Top sample                         & $\boldsymbol{k_2}$ (MPa)           &   80        & $\boldsymbol{k}$ (MPa)                &   470               \\
                         & $\boldsymbol{\eta_2}$ (MPa s)           &  300         &                 &                  \\
\midrule
\multirow{4}*{Collagen fibrils \cite{shen2011viscoelastic}}   & $\boldsymbol{k_0}$ (MPa) &  414         & $\boldsymbol{\eta}$ (MPa s)&  5100                 \\
                         & $\boldsymbol{k_1}$ (MPa)       &   78        & $\boldsymbol{c_\beta}$ (MPa s$^{\beta}$)                &  162                 \\
                         & $\boldsymbol{\eta_1}$ (MPa s)        & 331          & $\boldsymbol{\beta}$              &  0.21                 \\
Middle sample            & $\boldsymbol{k_2}$ (MPa)           &   62        & $\boldsymbol{k}$ (MPa)                 &   408               \\
                         & $\boldsymbol{\eta_2}$ (MPa s)           &  2700         &                 &                  \\
\midrule
\multirow{4}*{Collagen fibrils \cite{shen2011viscoelastic}}   & $\boldsymbol{k_0}$ (MPa) &  395 & $\boldsymbol{\eta}$ (MPa s) &  4400   \\
                         & $\boldsymbol{k_1}$ (MPa)       &   75       & $\boldsymbol{c_\beta}$ (MPa s$^{\beta}$)                &  120                 \\
                         & $\boldsymbol{\eta_1}$ (MPa s)        & 3080          & $\boldsymbol{\beta}$              &  0.12                 \\
{Bottom sample}  & $\boldsymbol{k_2}$ (MPa)            &   41        & $\boldsymbol{k}$ (MPa)                 &   390             \\
                         & $\boldsymbol{\eta_2}$ (MPa s)           &  163         &                 &                  \\
\bottomrule
\end{tabular*}
\label{tab:3}
\end{table}

\begin{table}[t]
\caption{Fitted parameters fractional model from figure \ref{fig:f16}}
\centering
\begin{tabular*}{0.6\textwidth}{l@{\extracolsep{\fill}}cccc}
\toprule
\textbf{t-t$_c$} & $\boldsymbol{\eta}$ (Pa) & $\boldsymbol{c}_{\boldsymbol{\beta}}$ (Pa s$^\beta$) & $\boldsymbol{\beta}$ & $\boldsymbol{k}$ (Pa) \\
\midrule
-6 &  3.9  & 9.4  & 0.71  & 0.0  \\
\midrule
-2 &  52.3  & 50.5  & 0.61  & 0.0  \\
\midrule
0&  336.3  & 121.0  & 0.54  & 20.2  \\
\midrule
2 &  1122.7  & 225.2  & 0.46  & 101.3  \\
\midrule
6 &  2963.5  & 920.6  & 0.36  & 1923.0  \\
\bottomrule
\end{tabular*}
\label{tab:4}
\end{table}

\begin{figure}[!h]
  \centering
 \includegraphics[width=0.6\textwidth]{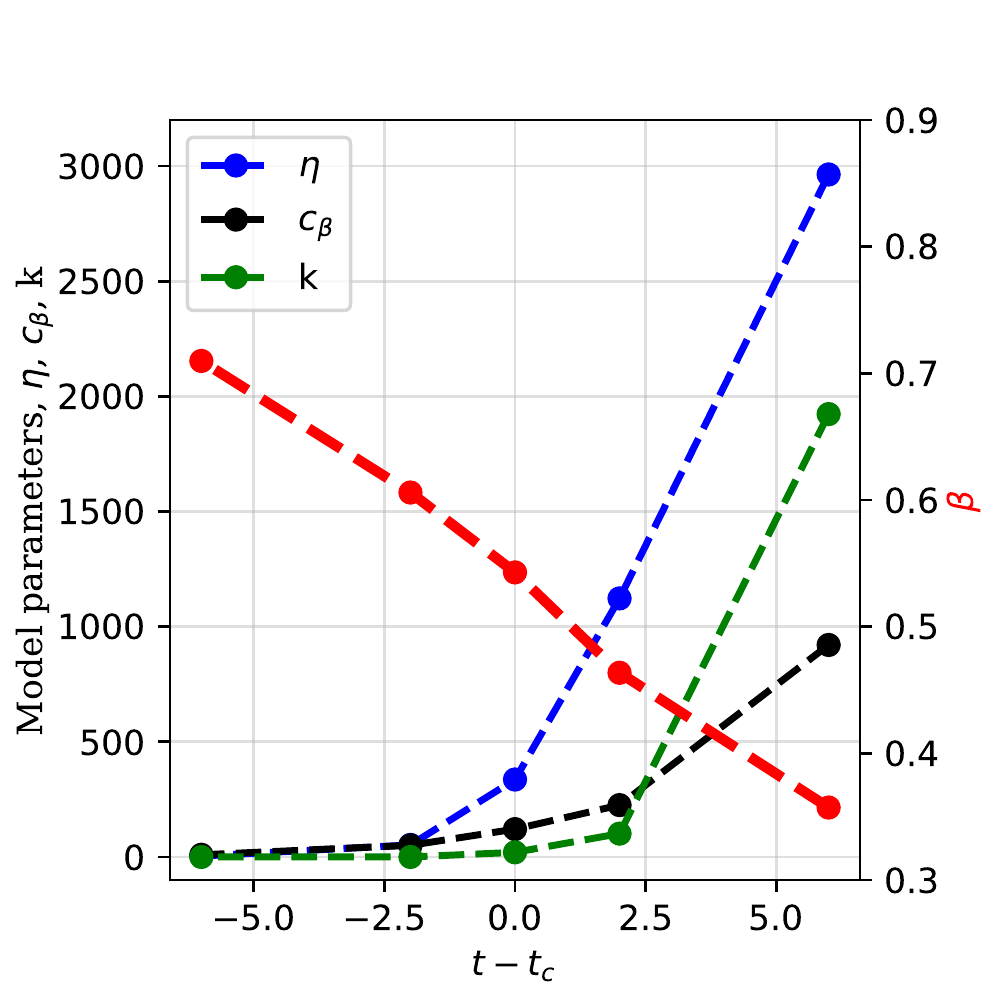}
  \caption{Fitted parameters fractional model from figure \ref{fig:f16}.}
  \label{fig:params}
\end{figure}

\printbibliography

\end{document}